\documentclass[fleqn,usenatbib]{mnras}

\usepackage[T1]{fontenc}

\DeclareRobustCommand{\VAN}[3]{#2}
\let\VANthebibliography\thebibliography
\def\thebibliography{\DeclareRobustCommand{\VAN}[3]{##3}\VANthebibliography}

\usepackage{graphicx}	
\usepackage{amsmath}	
\usepackage{amssymb}	
\usepackage{multicol}        
\usepackage{bm}		
\usepackage{pdflscape}	
\usepackage{hyperref}	
\usepackage[usenames,dvipsnames]{xcolor} 

\usepackage{newtxtext,newtxmath}

\def\gsim{\rlap{\raise0.7pt\hbox{$>$}}{\lower 0.8ex\hbox{$\sim$}}}
\def\lsim{\mathrel{\rlap{\lower3.0pt\hbox{$\sim$}}
            \raise0.7pt\hbox{$<$}}}

\title[Parameters constraint via \textsc{camels} and observations]{CASCO: Cosmological and AStrophysical parameters from Cosmological simulations and Observations -- I.  Constraining physical processes in local star-forming galaxies}

\author[V. Busillo et al.]{
V. Busillo,$^{1,2,3}$\thanks{E-mail: valerio.busillo@inaf.it}
C. Tortora,$^{2}$\thanks{E-mail: crescenzo.tortora@inaf.it}
N. R. Napolitano,$^{1,4}$
L. V. E. Koopmans,$^{5}$
G. Covone,$^{1,2,3}$
F. Gentile,$^{6,7}$
\and
L. K. Hunt.$^{8}$
\\
$^{1}$Dipartimento di Fisica “E. Pancini”, Universit\`{a} degli studi di Napoli Federico II, Compl. Univ. di Monte S. Angelo, Via Cintia, I-80126 Napoli, Italy\\
$^{2}$INAF -- Osservatorio Astronomico di Capodimonte, Salita Moiariello 16, I-80131, Napoli, Italy\\
$^{3}$INFN, Sez. di Napoli, Compl. Univ. di Monte S. Angelo, Via Cintia, I-80126 Napoli, Italy\\
$^{4}$School of Physics and Astronomy, Sun Yat-sen University Zhuhai Campus, 2 Daxue Road, Tangjia, Zhuhai, Guangdong 519082, PR China\\
$^{5}$Kapteyn Astronomical Institute, University of Groningen, P.O Box 800, 9700 AV Groningen, The Netherlands\\
$^{6}$University of Bologna, Department of Physics and Astronomy (DIFA), Via Gobetti 93/2, I-40129, Bologna, Italy\\
$^{7}$INAF -- Osservatorio di Astrofisica e Scienza dello Spazio, via Gobetti 93/3 - 40129, Bologna - Italy\\
$^{8}$INAF -- Osservatorio Astrofisico di Arcetri, Largo E. Fermi, 5, 50125, Firenze, Italy
}

\date{Accepted XXX. Received YYY; in original form ZZZ}

\pubyear{2023}

\begin{document}
\label{firstpage}
\pagerange{\pageref{firstpage}--\pageref{lastpage}}
\maketitle

\begin{abstract}
We compare the structural properties and dark matter content of star-forming galaxies taken from the \textsc{camels} cosmological simulations to the observed trends derived from the SPARC sample in the stellar mass range $[10^{9}, 10^{11}]\,\textrm{M}_{\odot}$, to provide constraints on the value of cosmological and astrophysical (SN- and AGN-related) parameters. We consider the size-, internal DM fraction-, internal DM mass- and total-stellar mass relations for all the 1065 simulations, all having different cosmological and astrophysical parameters, from the IllustrisTNG, SIMBA and ASTRID suites of \textsc{camels}, and search for the parameters that minimize the $\chi^{2}$ with respect to the observations. For the IllustrisTNG suite, we find the following constraints for the cosmological parameters: $\Omega_{\textrm{m}} = 0.27_{-0.05}^{+0.01}$, $\sigma_{8} = 0.83_{-0.11}^{+0.08}$ and $S_{8} = 0.78_{-0.09}^{+0.03}$, which are consistent within $1\sigma$ with the results from the nine-year WMAP observations. SN feedback-related astrophysical parameters, which describe the departure of outflow wind energy per unit star formation rate and wind velocity from the reference IllustrisTNG simulations, assume the following values: $A_{\textrm{SN1}} = 0.48_{-0.16}^{+0.25}$ and $A_{\textrm{SN2}} = 1.21_{-0.34}^{+0.03}$, respectively. Therefore, simulations with a lower value of outflow wind energy per unit star formation rate with respect to the reference illustrisTNG simulation better reproduce the observations. Variation of AGN feedback parameters, on the other hand, show negligible effects on the scaling relation trends in the mass range probed. Simulations based on SIMBA and ASTRID suites predict central dark matter masses substantially larger than those observed in real galaxies, which can be reconciled with observations only by requiring values of $\Omega_{\textrm{m}}$ inconsistent with cosmological constraints for SIMBA, or simulations characterized by unrealistic galaxy mass distributions for ASTRID.
\end{abstract}




\section{Introduction}

In the $\Lambda$CDM paradigm of structure formation, the large scale structure of the Universe (LSS) originates from tiny random fluctuations of the primordial dark matter density field, which are suppressed or grow according to various properties of these primordial overdensities, such as their scale. These growing fluctuations of dark matter then collapse from the ambient background, and start accreting primordial gas within them, sparking the formation of the primordial galaxies. In a `bottom-up' scenario of galactic formation, these primordial objects then start to merge with one another under the influence of gravity, gradually forming the most massive structures of the universe, such as galaxies and cluster of galaxies (\citealt{Springel2001}).

Scaling relations are a result of the physics behind galaxy formation and evolution: if gravity is the predominant process, then theoretical models predict simple scaling relations between various basilar halo properties, such as the Tully-Fisher relation \citep{Tully1977}, which relates the rotational velocity of spiral galaxies, $V_{\textrm{max}}$, to their intrinsic luminosity, $L$, which is itself proportional to mass (baryonic Tully-Fisher relation, \citealt{McGaugh2000}); the Faber-Jackson relation \citep{Faber1976}, which relates the central velocity dispersion of passive galaxies with their intrinsic luminosity; and the Fundamental Plane \citep{Djorgovski1987}, a three-dimensional manifold which relates effective radius $R_{\textrm{e}}$, mean surface brightness at $r=R_{\textrm{e}}$ and central velocity dispersion, $\sigma_{\textrm{c}}$, for passive galaxies.

There is general consensus \citep{McNamara2007, Dutton2009} that secondary, baryonic processes, such as active galactive nuclei (AGN) and supernovae (SN) feedback, need to be included in order to correctly reproduce the observed relations between galaxy parameters. Outflows driven by stellar winds and SN explosions are expected to dominate in the lower-mass regime, while at higher masses, outflows tend to be powered by feedback from active galactic nuclei \citep{Tremonti2004, Zahid2014, Tortora2019, Lara2019}.
Galactic winds generated by stars and SN, for example, are regulating the baryon cycle, e.g. the star formation and the metallicity in the interstellar medium \citep{Tortora2022}, shaping the “main sequence” correlation between $M_{*}$ and the star-formation rate (SFR, \citealt{Brinchmann2004}) and the mass–metallicity relation (MZR, \citealt{Tremonti2004}). AGN feedback is instead required in massive galaxies to efficiently quench the star formation and make these galaxies passive \citep{Lagos2008}. Various studies have reported deviations for low-mass systems from the trends expected from simple models in which gravity is the only dominant process: these deviations could indicate that non-gravitational processes may significantly impact the evolution of these systems \citep{Gastaldello2007, Sun2009, Eckmiller2011}. What is the relative contribution of these processes, however, is still debated.

These findings have triggered a renewed interest in turning to cosmological simulations, to try and take such processes into account, for example succeeding in simulating the feedback between the central super-massive black hole (SMBH) of a galaxy and its global properties \citep{Puchwein2008}. In this context, simulations prove to be very useful tools. They can be used for instance in conjunction to machine learning algorithms to predict galaxy properties via the use of scaling relations \citep{Shao2022b}, and the best cosmological parameter combinations given the physical properties of a sample of galaxy clusters \citep{Qiu2023} or for single galaxies as in \cite{Villaescusa2022} and \cite{Echeverri2023}, or to determine the effects of feedback mechanisms on the morphology of galaxies \citep{Okamoto2005}, on the relation between total mass density profile and dark matter fraction within the half-mass radius of galaxies \citep{Remus2017}, on the generation of galactic winds \citep{Hopkins2012} and on structural and dynamical properties of galaxies \citep{Irodotou2022}.

The variation of scaling relation trends with the underlying cosmology and astrophysical recipes proves to be a promising tool for cosmological tests based on simulations, in that cosmological parameters of simulations can be easily modified. Work on comparing observations to simulated data has been performed on the concentration-mass relation \citep{Shao2022a}, the baryonic Tully-Fisher relation \citep{Goddy2023} and dark matter fraction and mass density slope in massive galaxies \citep{Mukherjee2018, Mukherjee2021, Mukherjee2022}. But, to our knowledge, there is still no attempt to use these comparisons as a tool to constrain cosmology or astrophysics in the required fine details. This is because simulations such as IllustrisTNG often assume a fixed cosmology and sub-grid parameters, and are also calibrated on some observed relations. The "Cosmology and Astrophysics with MachinE Learning Simulations" (\textsc{camels}, \citealt{Villaescusa2021}) cosmological simulations provide for the first time the chance to investigate the impact of a wide range of cosmologies and physical processes on observed scaling relations. the \textsc{camels} simulations do not fix the values of the cosmological and astrophysical parameters, but vary them very finely without requiring any calibration with the observations (except for one of them, the fiducial cosmology). This is done because \textsc{camels} is mainly used to train machine learning algorithms for predicting a certain set of cosmological and astrophysical parameter set from the observations, and as such it is perfectly suited also for standard statistical analysis such as the one proposed in this work.

In this paper, we present the project {\it CASCO: Cosmological and AStrophysical parameters from Cosmological simulations and Observations}. We start testing the predictive power
encoded in various scaling relations of star-forming galaxies, by comparing \textsc{camels} simulations to observed trends inferred from the Spitzer Photometry \& Accurate Rotation Curves (SPARC, \citealt{Lelli2016}) sample, to constrain cosmological and astrophysical parameters (SN- and AGN-related ones). We demonstrate the potentiality of this method using half-mass radius and dark-matter-related quantities in local star-forming galaxies, planning to extend the analysis to other galaxy types, redshifts and galaxy parameters in future papers.

The paper is organized as follows: in Section \ref{sec:data_CAMELS_SPARC}, we present an overview of the \textsc{camels} simulations, the selection criteria of the simulated galaxies that we will consider in the analysis and the sample of observed SPARC galaxies that we considered for comparison with the simulated data. In Section \ref{sec:Analysis}, we compare the scaling relations observed in the simulated data to the respective observed trends, and give constraints for the cosmological and astrophysical parameters. We provide a physical interpretation of the results in Section \ref{sec:Discussion}, and give our conclusions in Section \ref{sec:Conclusions}.

\section{Observations and \textsc{camels} simulations}\label{sec:data_CAMELS_SPARC}
In this section, we describe the data samples used in this paper. in Section \ref{sec:SPARC_data} we introduce the SPARC sample, a catalog of local star-forming galaxies, while in Section \ref{sec:CAMELS_simulations} we introduce the \textsc{camels} cosmological simulations.

\subsection{SPARC data}\label{sec:SPARC_data}
The observational data used in the analysis come from the sample of 175 disc galaxies with near-infrared photometry and H \textsc{i} rotation curves (SPARC, \citealt{Lelli2016}). This sample is neither statistically complete, nor volume-limited, but it is nevertheless representative of the population of disc galaxies in the local Universe. The SPARC sample's morphological types range from irregular (Im/BCD) to lenticular (S0), and cover a large range of effective radii ($\sim 0.3$ to $\sim 15\;\textrm{kpc}$), rotation velocities ($\sim 20$ to $\sim 300\;\textrm{km}\,\textrm{s}^{-1}$) and gas contents ($\sim 0.01$ to $\sim 10 \; M_{\textrm{HI}} / L_{[3.6\,\mu\textrm{m}]}/(\textrm{M}_{\odot} / \textrm{L}_{\odot})$). The radial velocity curves $v(r)$ for the galaxies have been obtained based primarily on HI measurements. Total mass enclosed within a sphere of radius $r$ is determined via the radial velocity curves, using the formula $M(r)=v^2 r/G$. Total stellar mass is obtained from the total luminosity assuming a constant stellar mass-to-light ratio at $3.6\;\mu\textrm{m}$ equal to $\Upsilon_{*} = 0.6\,\Upsilon_{\odot}$ (for details, see \citealt{Tortora2019}). Given that we are comparing these data with simulations, for which only tridimensional structural quantities are available, we cannot perform a comparison by using directly the effective radius, which is a projected quantity. We thus converted the SPARC galaxies' effective radii into stellar half-mass radii, $R_{*,1/2}$, by multiplying the respective effective radii by a constant factor of $\sim 1.35$ \citep{Wolf2010}. For a discussion on the impact of fixing $\Upsilon_{*}$ to $0.6\,\Upsilon_\odot$ and of converting between projected and 3D radii, see Appendix \ref{sec:const_ML_ratio_and_2D_3D_conversion}.

Following \cite{Tortora2019}, of the 175 galaxies in the SPARC sample we consider only those with inclinations larger than 30°, because rotation velocities for face-on systems are highly uncertain. This procedure does not introduce a selection bias, because the galaxies' orientation in the sky is random. We also omit from the final sample those galaxies for which the effective radius $R_{\textrm{e}}$ is not covered by the rotation curve, in order to avoid extrapolations. 

The final sample thus consists of 152 galaxies, for which we consider the total stellar mass ($M_{*}$), the stellar half-mass radius ($R_{*,1/2}$) and the total, stellar and gas mass within the stellar half-mass radius ($M_{1/2}$, $M_{*,1/2}$ and $M_{\textrm{g},1/2}$, respectively).  The dark matter mass within the stellar half-mass radius, $M_{\textrm{DM},1/2}$, is obtained by subtracting the stellar and gas mass contributes from $M_{1/2}$. Total (virial) masses are taken from \cite{Posti2019}, obtained by modelling the rotation curves with a baryonic component plus a Navarro-Frenk-White \citep{Navarro1996} model for DM. None of the observables depend directly on the cosmological parameters because SPARC is a catalog of local galaxies, for which distances are measured with direct methods.

\subsection{\textsc{camels} simulations} \label{sec:CAMELS_simulations}
The simulated galaxy data come from \textsc{camels}, a suite of 6325 cosmological simulations of an Universe volume equal to $25\;h^{-1}\,\textrm{Mpc}$ \citep{Villaescusa2021}. Approximately half of these are gravity-only N-body simulations, while the other half are hydrodynamical simulations, which are obtained by implementing three different hydrodynamical sub-grid models: IllustrisTNG \citep{Pillepich2018}, SIMBA \citep{Davé2019} and ASTRID \citep{Bird2022, Ni2022}.

The mass resolution for the dark matter (DM) particles is $M_{\textrm{DM, min}} = 6.49\times 10^{7} (\Omega_{\textrm{m}}-\Omega_{\textrm{b}})/0.251\;h^{-1}\,\textrm{M}_{\odot}$ (which, for a simulation having $\Omega_{\textrm{m}}=0.3$, is equal to $9.67\times 10^{7}\; \textrm{M}_{\odot}$), while for the gas particles is $M_{\textrm{g, min}} = 1.89\times 10^{7}\; \textrm{M}_{\odot}$. These values are the same for all \textsc{camels} suites. Galaxies/subhalos are identified using the \textsc{subfind} subhalo finder algorithm \citep{Springel2001}. In \textsc{camels}, the following cosmological parameters are fixed:  $\Omega_{\textrm{b}} = 0.049$, $\Omega_{\textrm{k}} = 0$, $n_{\textrm{s}} = 0.9624$, $h=0.6711$ and $M_{\nu} = 0.0\;\textrm{eV}$, where $h = H_{0}/100\;\textrm{km}\,\textrm{s}^{-1}\,\textrm{Mpc}^{-1}$. The assumed equation of state of dark energy is $P(\rho)=w\rho$, with $w=-1$. The values of the matter density parameter, $\Omega_{\textrm{m}}$, and of the amplitude of the linear matter density fluctuations, $\sigma_{8}$, are instead free parameters that depend on the particular simulation considered. For all \textsc{camels} simulation suites, a \cite{Chabrier2003} initial mass function (IMF) is assumed.

As detailed in \cite{Villaescusa2021}, for each simulation six parameters are varied: two cosmological parameters ($\Omega_{\textrm{m}}$, $\sigma_{8}$) and four astrophysical parameters ($A_{\textrm{SN1}}$, $A_{\textrm{AGN1}}$, $A_{\textrm{SN2}}$ and $A_{\textrm{AGN2}}$), each related to a different astrophysical process. In particular, $A_{\textrm{SN1}}$ and $A_{\textrm{SN2}}$ are related to the supernovae feedback mechanisms, while $A_{\textrm{AGN1}}$ and $A_{\textrm{AGN2}}$ are related to AGN feedback. It should be noted that the astrophysical parameters have different physical meanings for each of the three suites, and should be considered as completely different parameters. As such, from here on, we will refer to the four astrophysical parameters associated to the SIMBA simulations with a tilde (e.g. $\tilde{A}_{\textrm{SN1}}$) and to those of ASTRID with a hat (e.g. $\hat{A}_{\textrm{SN1}}$), in order to avoid confusion.

For our analysis, we used all three of the hydrodynamical simulation suites, whose specific properties will be discussed more in detail in the following sections.

\subsubsection{IllustrisTNG suite}
IllustrisTNG utilizes the \textsc{arepo} code \citep{Springel2010} to solve the coupled gravity and magneto-hydrodynamics equations for each particle, in addition to sub-grid physics models for astrophysical processes such as star-formation, supernovae feedback, growth of supermassive black holes and AGN feedback. The gravitational softening length for dark matter is equal to $\epsilon_{\textrm{min}} = 2\;\textrm{kpc}$ comoving.

In the IllustrisTNG suite, the $A_{\textrm{SN1}}$ and $A_{\textrm{SN2}}$ parameters both contribute to the wind mass loading factor at injection, $\eta_{\textrm{w}} := \dot{M_{\textrm{g}}}/\dot{M}_{\textrm{SFR}}$, where $\dot{M}_{\textrm{g}}$ is the rate of gas mass inside a galaxy converted into ejected wind mass, and $\dot{M}_{\textrm{SFR}}$ is the local instantaneous star formation rate. This is an important parameter for describing the effects of galactic winds on the chemical evolution of galaxies, because the wind mass loading characterizes the dominance of bulk outflows over gas accretion, and comes into play in the equilibrium condition between inflows and outflows for a galaxy \citep{Tortora2022}.

Following \cite{Pillepich2018}, the wind mass loading in IllustrisTNG can be written as:

\begin{equation}
\eta_{\textrm{w}}=\frac{2}{v_{\textrm{w}}^{2}}e_{\textrm{w}}\left(1-\tau_{\textrm{w}}\right),
\end{equation}
\\
where $\tau_{\textrm{w}}=0.1$ is the thermal fraction, $e_{\textrm{w}}$ is the galactic wind energy per unit star formation rate, written as:

\begin{align}
e_{\textrm{w}}= &A_{\textrm{SN1}} \times \overline{e}_{\textrm{w}}\left[f_{\textrm{w},Z}+\frac{1-f_{\textrm{w},Z}}{1+(Z/Z_{w,\textrm{ref}})^{\gamma_{\textrm{w},Z}}}\right]\nonumber \\
 & \times N_{\textrm{SNII}}E_{\textrm{SNII,51}}\times10^{51}\;\textrm{erg}\,\textrm{M}_{\odot}^{-1},\label{eq:IllustrisTNG_energy_unit_SFR}
\end{align}
\\
with $Z$ metallicity of gas cells,  $\overline{e}_{\textrm{w}}$ wind energy factor, $f_{\textrm{w},Z}$ $Z$-dependence reduction factor, $Z_{\textrm{w, ref}}$ $Z$-dependence reference metallicity, $\gamma_{\textrm{w},Z}$ $Z$-dependence reduction power, $N_{\textrm{SNII}}$ number of SNII per formed stellar mass and $E_{\textrm{SNII,51}}$ available energy per core-collapse SNe in units of $10^{51}\;\textrm{erg}$, as reported in \cite{Pillepich2018}, while $v_{\textrm{w}}$ is the galactic wind speed at injection, given by\footnote{Notice that we modified equation \eqref{eq:IllustrisTNG_wind_velocity} with respect to the version reported in \cite{Pillepich2018}, according to \cite{Ni2023} (Appendix A1, footnote 4).}:

\begin{equation}
v_{\textrm{w}}=\textrm{max}\left[A_{\textrm{SN2}}\,\kappa_{\textrm{w}}\,\sigma_{\textrm{DM}}\left(\frac{H_{0}}{H(z)}\right)^{1/3},\;v_{\textrm{w, min}}\right],\label{eq:IllustrisTNG_wind_velocity}
\end{equation}
\\
where $\kappa_{\textrm{w}}$ is the wind velocity factor, also reported in \cite{Pillepich2018}, $\sigma_{\textrm{DM}}$ is the 1D local dark matter velocity dispersion and $v_{\textrm{w, min}} = 350\;\textrm{km}\,\textrm{s}^{-1}$ is the wind velocity floor at injection.

The AGN feedback parameters, instead, modulate the low accretion rate kinetic SMBH feedback mode, with $A_{\textrm{AGN1}}$ influencing the power injected in the kinetic mode:

\begin{equation}
\dot{E}_{\textrm{low}}=A_{\textrm{AGN1}}\times\textrm{min}\left[\frac{\rho}{0.05\,\rho_{\textrm{SF, thresh}}},\;0.2\right]\times\dot{M}_{\textrm{BH}}c^{2}, \label{eq:IllustrisTNG_power_injected}
\end{equation}
\\
where $\rho$ is the gas density around the SMBH, $\rho_{\textrm{SF, thresh}}$ is the density threshold for star formation and $\dot{M}_{\textrm{BH}}$ is the accretion rate of the central galactic supermassive black hole, while $A_{\textrm{AGN2}}$ influences the `burstiness' of the central black hole, that is, the rate at which the supermassive black hole ejects energy, which happens every time the accreted energy equals the following threshold value:

\begin{equation}
E_{\textrm{inj, min}}=A_{\textrm{AGN2}}\times f_{\textrm{re}}\times\frac{1}{2}m_{\textrm{enc}}\tilde{\sigma}_{\textrm{DM}}^{2}, \label{eq:IllustrisTNG_energy_threshold}
\end{equation}
\\
where $f_{\textrm{re}}=20$ is a constant of the fiducial TNG model, $\tilde{\sigma}_{\textrm{DM}}$ is the one-dimensional dark matter velocity dispersion around the central SMBH, and $m_{\textrm{enc}}$ is the gas mass inside the feedback sphere.

\subsubsection{SIMBA suite}
SIMBA relies on the \textsc{gizmo} \citep{Hopkins2015} code for solving the equations, in its `Meshless Finite Mass' (MFM) mode. The gravitational softening length in SIMBA is an adaptive parameter: as a conservative choice, we decided to consider a fixed minimum gravitational softening length of $\epsilon_{\textrm{min}} = 0.75\;\textrm{kpc}$ comoving (see \citealt{Davé2019}, Table 1 for details).

In the SIMBA suite, the wind mass loading factor is directly regulated only by the parameter $\tilde{A}_{\textrm{SN1}}$, via a power law fit based on \cite{Alcazar2017} FIRE `zoom-in' simulations:

\begin{equation}
\eta_{\textrm{w}}=\tilde{A}_{\textrm{SN1}}\times\begin{cases}
\displaystyle9\left(\frac{M_{*}}{M_{0}}\right)^{-0.317} & \textrm{, if }M_{*}<M_{0},\\
\displaystyle9\left(\frac{M_{*}}{M_{0}}\right)^{-0.761} & \textrm{, if }M_{*}>M_{0},
\end{cases}
\label{eq:SIMBA_wind_mass_loading_trend}
\end{equation}
\\
where $M_{0} = 5.2\times 10^{9}\;\textrm{M}_{\odot}$. It should be noted that the wind mass loading trend of \cite{Alcazar2017} differs from the one used in \cite{Muratov2015}, which is also based on the FIRE simulations, in that the former tracks individual particles in order to quantify the mass outflow rates out of the star-forming region, while the latter computes outflow rates based on mass advection across a boundary at one quarter of the virial radius. The consequence of this is that the slope of the relation in \cite{Alcazar2017} is similar to the one in \cite{Muratov2015}, but the former shows roughly double the amplitude of the latter, and is much steeper above $M_{0}$. The parameter $\tilde{A}_{\textrm{SN2}}$, instead, regulates the outflow wind velocity as a function of the circular velocity \citep{Muratov2015}:

\begin{equation}
v_{\textrm{w}}=\tilde{A}_{\textrm{SN2}}\times1.6\left(\frac{v_{\textrm{circ}}}{200\;\textrm{km}\,\textrm{s}^{-1}}\right)^{0.12}v_{\textrm{circ}}+\Delta v(0.25R_{\textrm{vir}}), \label{eq:SIMBA_wind_velocity}
\end{equation}
\\
where $\Delta v (0.25 R_{\textrm{vir}})$ is the velocity corresponding to the potential difference between the launch point and $0.25 R_{\textrm{vir}}$.
Finally, the AGN feedback parameters for the SIMBA suite regulate the total momentum flux of the ejected gas, in the form of relativistic jets, via the relation:

\begin{equation}
\dot{P}_{\textrm{out}}\equiv\dot{M}_{\textrm{out}}v_{\textrm{out}}=\tilde{A}_{\textrm{AGN1}}\times20\,L_{\textrm{bol}}/c, \label{eq:SIMBA_total_momentum_flux}
\end{equation}
\\
where $L_{\textrm{bol}}=\epsilon_{\textrm{r}} \dot{M}_{\textrm{BH}} c^2$ is the bolometric luminosity and $\epsilon_{\textrm{r}}=0.1$ is the radiative efficiency, and the outflow velocity of the SMBH jet emissions:

\begin{equation}
v_{\textrm{out}}=\begin{cases}
\displaystyle v_{\textrm{rad}}+\tilde{A}_{\textrm{AGN2}}\times v_{\textrm{jet}} & \textrm{, if }\begin{cases}
\lambda_{\textrm{Edd}}<0.2\\
M_{\textrm{BH}}>10^{7.5}\,\textrm{M}_{\odot}
\end{cases},\\
\displaystyle v_{\textrm{rad}} & \textrm{, otherwise.} \label{eq:SIMBA_outflow_velocity}
\end{cases}
\end{equation}

\subsubsection{ASTRID suite}
ASTRID uses a new version of the \textsc{mp-gadget} code, a modified version of \textsc{gadget-3} \citep{Springel2005}, to solve gravity with an $N$-body tree-particle-mesh (TreePM) approach, hydrodynamics with smoothed particle hydrodynamics (SPH) method and astrophysical processes with a series of subgrid models. The gravitational softening length in ASTRID is $\epsilon_{\textrm{min}} = 2.2\;\textrm{kpc}$ comoving.

In ASTRID, the parameters $\hat{A}_{\textrm{SN1}}$ and $\hat{A}_{\textrm{SN2}}$ have a similar role to the SIMBA parameters $\tilde{A}_{\textrm{SN1}}$ and $\tilde{A}_{\textrm{SN2}}$, but the formula for the wind mass loading is different: in the case of ASTRID, $\hat{A}_{\textrm{SN1}}$ directly controls the wind mass loading, but via the formula:

\begin{equation}
\eta_{\textrm{w}} = \hat{A}_{\textrm{SN1}}\times \left(\frac{\sigma_{0,\textrm{fid}}}{v_{\textrm{w}}}\right)^{2}, \label{eq:ASTRID_wind_mass_loading_trend}
\end{equation}
\\
where $\sigma_{0,\textrm{fid}} = 353\;\textrm{km/s}$ \citep{Bird2022}. The parameter $\hat{A}_{\textrm{SN2}}$, similarly to $A_{\textrm{SN1}}$ in equation \eqref{eq:IllustrisTNG_wind_velocity} instead, regulates the wind velocity through the following formula:

\begin{equation}
v_{\textrm{w}} = \hat{A}_{\textrm{SN2}} \times \kappa_{\textrm{w}}\,\sigma_{\textrm{DM}}, \label{eq:ASTRID_wind_velocity}
\end{equation}
\\
where $\kappa_{\textrm{w}} = 3.7$ and $\sigma_{\textrm{DM}}$ is the same as in equation \eqref{eq:IllustrisTNG_wind_velocity}, following the IllustrisTNG model, but without the wind velocity floor at injection and the dependency on $H(z)$. 

For the AGN feedback in ASTRID, the parameters $\hat{A}_{\textrm{AGN1}}$ and $\hat{A}_{\textrm{AGN2}}$ regulate the kinetic and thermal feedback modes, respectively, via the following equations:

\begin{equation}
\begin{cases}
\displaystyle \Delta\dot{E}_{\textrm{low}} = \hat{A}_{\textrm{AGN1}}\times\epsilon_{\textrm{f,kin}}\,\dot{M}_{\textrm{BH}}c^2 & ,\,\lambda_{\textrm{Edd}}<\chi_{\textrm{thr}},\\
\displaystyle \Delta\dot{E}_{\textrm{high}} = \hat{A}_{\textrm{AGN2}}\times \epsilon_{\textrm{f,th}}\epsilon_{\textrm{r}}\,\dot{M}_{\textrm{BH}}c^2 & ,\,\lambda_{\textrm{Edd}}>\chi_{\textrm{thr}}, \label{eq:ASTRID_AGN_feedback}
\end{cases}
\end{equation}
\\
where $\Delta\dot{E}_{\textrm{low}}$ and $\Delta\dot{E}_{\textrm{high}}$ are respectively the power injected in the kinetic and thermal mode, $\chi_{\textrm{thr}}$ is the Eddington threshold, $\epsilon_{\textrm{r}}$ is the mass-to-light conversion efficiency and $\epsilon_{\textrm{f,kin}}$ and $\epsilon_{\textrm{f,th}}$ are the fraction of the radiation energy kinetically and thermally injected into the surrounding gas, respectively (for more details on the values assumed by these parameters, see \citealt{Ni2023}).

\subsubsection{General considerations about the \textsc{camels} fiducial simulations}
It should be noted that, in equations (\ref{eq:IllustrisTNG_energy_unit_SFR}-\ref{eq:ASTRID_AGN_feedback}), an unitary value of the astrophysical parameters implies that the equations reduce exactly to the relations reported in \cite{Pillepich2018}, \cite{Weinberger2017}, \cite{Alcazar2017}, \cite{Muratov2015}, \cite{Bird2022} and \cite{Ni2022}. These relations are the ones that have been implemented in the `original' IllustrisTNG, SIMBA and ASTRID simulation runs, as detailed in \cite{Nelson2019}, \cite{Davé2019}, \cite{Bird2022} and \cite{Ni2022}. We will thus refer to simulations with unit values of the astrophysical parameters and cosmological parameters equal to $\Omega_{\textrm{m}} = 0.30$ and $\sigma_{8} = 0.80$ as fiducial simulations.

\begin{figure*}

\centering
    \includegraphics[width=\linewidth]{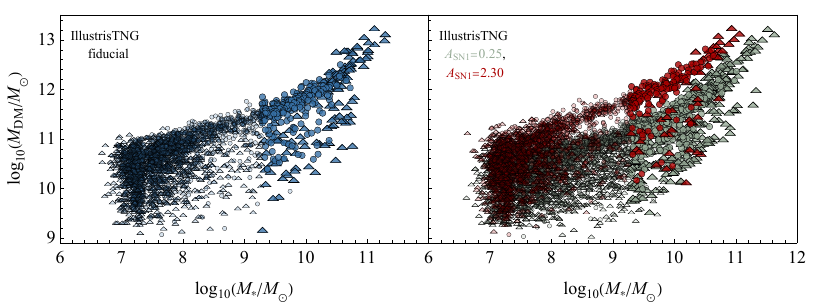}

\caption{\textit{Left panel}. Distribution of all galaxies from the IllustrisTNG fiducial simulation in the $M_{\textrm{DM}}$-$M_{*}$ plane. Opaque circles and triangles are star-forming and passive galaxies, respectively, which satisfy the filtering conditions detailed in Section \ref{sec:CAMELS_simulations}. Transparent circles and triangles are all the star-forming and passive objects which are under the filtering threshold. \textit{Right panel}.  Same as the left panel, but showing galaxies with all the parameters fixed to those of the reference simulations except for $A_{\textrm{SN1}}$, set to  $0.25$  (`$\textrm{1P\_3\_n5}$' simulation, dark green points) and $2.3$  (`$\textrm{1P\_3\_3}$' simulation, red points).}
\label{fig:selection_effects_sims}

\end{figure*}

It is important to note that, in \textsc{camels}, only the fiducial simulations have been calibrated to reproduce several galaxy properties. For the IllustrisTNG suite, the calibrations have been performed by using the galaxy stellar mass function, the stellar-to-halo mass relation, the total gas mass content within the virial radius $r_{500}$ of massive groups, the stellar mass-stellar size and the black hole mass - galaxy mass relations, all at $z=0$, and, finally, the functional shape of the cosmic star formation rate density for $z\lsim 10$ \citep{Pillepich2018}. For the SIMBA suite, the calibrations are based only on the stellar mass function, the cosmic SFR density and the black hole mass - galaxy mass relation \citep{Davé2019}. For the ASTRID suite, the free parameters of the UV-band dust optical depth have been calibrated against the observed galaxy UV luminosity function at redshift $z=4$, and applied to all redshifts \citep{Bird2022}. In all the other simulations, the subgrid parameters and the cosmological parameter values are varied without requiring the simulations to reproduce any kind of observation.

\subsubsection{Simulation types and physical quantities used}
The three suites contain four different varieties of simulations. These include:
\begin{itemize}
\item 27 fiducial simulations for which only the seed for generating the initial conditions is varied (cosmic variance set, CV);
\item 61 simulations in which the value of the cosmological and astrophysical parameters is varied one at a time, with a fixed seed value for all the simulations (1-parameter set, 1P). In particular, the simulation `1P\_1\_0' is a fiducial simulation, used for reference;
\item  1000 simulations in which the value of the cosmological and astrophysical parameters, as well as the seed value, are varied randomly, by using Latin-hypercube sampling (Latin-hypercube set, LH);
\item 4 simulations in which the cosmological parameters and the seed value are fixed, and the astrophysical values are set to extreme values (extreme set, EX), such as very efficient supernova feedback ($A_{\textrm{SN1}} = \tilde{A}_{\textrm{SN1}} = 100.00$), very efficient AGN feedback ($A_{\textrm{AGN1}} = \tilde{A}_{\textrm{AGN1}} = 100.00$) and no feedback (all astrophysical parameters equal to zero). In particular, the EX\_0 is a fiducial simulation, used for reference.
\end{itemize}

We made use of all these simulations both from the IllustrisTNG, SIMBA and ASTRID suites in our analysis. For the comparison with observations, we consider the following physical quantities:

\begin{itemize}
\item Stellar half-mass radius, $R_{*,1/2}$, defined as the radius containing half of the total stellar mass of the galaxy;
\item Total stellar mass, $M_{*}$, defined as the sum of the masses of all star particles bound to a certain subhalo, as detected by \textsc{subfind};
\item Total mass, $M_{\textrm{tot}}$, defined as the sum of the masses of all particles/cells of every type (stellar, dark matter, gas, black hole) bound to a certain subhalo;
\item Stellar/DM/gas/total mass within the half-mass radius, $M_{*,1/2}$, $M_{\textrm{DM},1/2}$, $M_{\textrm{g},1/2}$ and $M_{1/2}$, respectively, defined as the sum of the masses of particles of the respective types which are within a sphere with radius equal to the stellar half-mass radius of a certain subhalo;
\item DM fraction within the stellar half-mass radius, $f_{\textrm{DM}}(<R_{*,1/2})$, defined as the ratio $M_{\textrm{DM},1/2}/M_{1/2}$;
\item Number of star particles within the stellar half-mass radius, $N_{*,1/2}$, defined as the ratio $M_{*,1/2}/M_{\textrm{g, min}}$;
\item Star formation rate, $\textrm{SFR}$, defined as the sum of the star formation rates of all star-forming gas cells of a certain subhalo;
\item Maximum rotational velocity of the spherically-averaged rotation curve, $V_{\textrm{max}}$, where all particle types (gas, stars, DM and SMBHs) are considered for its determination;
\item one-dimensional total velocity dispersion, $\sigma$, defined as the 3D velocity dispersion of all the member particles/cells bound to a certain subhalo, divided by $\sqrt{3}$;
\item one-dimensional local dark matter velocity dispersion around a star particle, $\sigma_{\textrm{DM}}$, defined as the 1D velocity dispersion of all dark matter particles within the comoving radius of a sphere centered on a certain star particle, enclosing the nearest $64\pm 1$ dark matter particles;
\item mean one-dimensional dark matter velocity dispersion, $\overline{\sigma}_{\textrm{DM}}$, defined as the mean of the distribution formed by all the 1D local dark matter velocity dispersions, $\sigma_{\textrm{DM}}$, around each star particle of the subhalo;
\item Total gas metallicity, $Z$, defined in the IllustrisTNG suite as the mass-weighted average metallicity of the gas cells bound to a certain subhalo for all gas cells within a sphere with radius associated to the maximum rotational velocity of the velocity curve, $V_{\textrm{max}}$.
\end{itemize}

The quantities $V_{\textrm{max}}$,  $\sigma$, $\sigma_{\textrm{DM}}$, $\overline{\sigma}_{\textrm{DM}}$ and $Z$ in particular have only been used for the evaluation of the wind mass loading at injection, in the analysis detailed in Section \ref{sec:wind_mass_loading_analysis}. The quantities that we are using are all tabulated de-projected values, obtained via the files `fof\_subhalo\_tab\_033.hdf5' (for IllustrisTNG and SIMBA) and `fof\_subhalo\_tab\_090.hdf5' (for ASTRID), available on the CAMELS website, relative to the $z=0$ snapshot. In future articles, we will consider the single particles associated to each subhalo also to evaluate numerically the corresponding projected quantities.

\subsubsection{Observational realism and cosmic variance}
It has to be noted that galaxy quantities are determined in a different way in simulation and real data. For example, effective radius is measured as the radius encompassing half of the total 
$[3.6\,\mu\textrm{m}]$ luminosity in SPARC (these wavelengths probe quite well the mass of the galaxies) and deprojected using a constant multiplicative factor, while in the simulations, it is defined as the radius containing half of the total stellar mass. Observed total masses are derived in \cite{Posti2019}, fitting an analytical galaxy model to rotation curves, while in simulations bounded star/gas/DM particles/cells are considered. Stellar mass is calculated in SPARC by using $[3.6\,\mu\textrm{m}]$ luminosity and a constant mass-to-light ratio, while the treatment is obviously more complex in the simulations.  The inclusion of more observational realism in the simulated quantities is difficult to treat and is beyond the scope of this paper. Howerer, we believe that possible differences arising from homogenizing the definition of galaxy quantities will induce secondary contributions, which will not strongly affect the results presented in this work.

Both cosmological simulations and observations are sampling only a limited volume of the Universe and therefore are affected by cosmic variance. Cosmic variance can potentially impact the physical properties and scaling relations resulting from both simulations and observations. A discussion on how cosmic variance affects the fiducial simulations is described in detail in Appendix \ref{sec:cosmic_variance}, using the Cosmic Variance set. Results show that the effects of cosmic variance on the properties considered in this paper is of the order of $10^{-2}$ dex for all simulation suites, and thus negligible.

\subsubsection{Filtering procedure}
In this paper, for each simulation we consider a filtered subset of all the subhalos detected by the \textsc{subfind} algorithm. This is done because some of the objects detected by the algorithm are not actual galaxies, but disk fragments or other artifacts, while other objects are not well-resolved, having smaller dimensions than the gravitational softening length, or with too few star particles inside the half-mass radius.

The parameters on which we base this filtering are the half mass radius, $R_{*,1/2}$, the number of star particles within the stellar half-mass radius, $N_{*,1/2}$, and the DM fraction within the stellar half-mass radius, $f_{\textrm{DM}}(<R_{*,1/2})$. We consider for the analysis only subhalos which have $R_{*,1/2}>\epsilon_{\textrm{min}}$, $N_{*,1/2} > 50$ and $f_{\textrm{DM}}(<R_{*,1/2})>0$.

Because SPARC is a sample of star-forming galaxies, we also performed a selection with respect to the specific star formation rate ($\textrm{sSFR} := \textrm{SFR}/M_{*}$, where $\textrm{SFR}$ is the galaxy's star formation rate). Following \cite{Bisigello2020}, we considered as star-forming galaxies only those subhalos which possess $\log_{10}(\textrm{sSFR}/\textrm{yr}^{-1}) > -10.5$. The effects of fixing a specific $\textrm{sSFR}$ threshold on the scaling relations are negligible and are discussed in Appendix \ref{sec:fixed_sSFR_contribution}.

The effects of the selections are shown in Fig. \ref{fig:selection_effects_sims}  for both the fiducial IllustrisTNG simulation and for two IllustrisTNG simulations from the `1P' set. The overall effect of the selection criteria is approximately a vertical cut in stellar mass, with the threshold located at $\log_{10}(M_{*}/\textrm{M}_{\odot}) \approx 9.25$, consistent with cuts performed during the analysis of the stellar mass function in \cite{Villaescusa2021}.

Our selection is very conservative when compared to other works with \textsc{camels}, e.g. in \cite{Villaescusa2022}, where the selection is based on considering only subhalos with total number of star particles greater than 20. In their case, the selection criteria effect is a cut in $M_{*}$, with threshold approximately equal to $\log_{10}(M_{*}/M_{\odot}) \approx 8.35$.

\begin{table*}
\centering

\caption{Chi-squared ($\chi^{2}$) and normalized chi-squared ($\tilde{\chi}^{2}$) values for the scaling relations considered in Section \ref{sec:sim_base_comparison}, associated to the IllustrisTNG, SIMBA and ASTRID fiducial simulations. First and second columns show the values associated to IllustrisTNG, third and fourth columns show those associated to SIMBA and the last two show those associated to ASTRID}. The last row presents the cumulative results, which represent the sum of the chi-squared values relative to the four scaling relations.

\begin{tabular}{ccccccc}
\hline 
\hline
relation & $\chi^{2}$ & $\tilde{\chi}^{2}$  & $\chi^{2}$  & $\tilde{\chi}^{2}$ & $\chi^{2}$ & $\tilde{\chi}^{2}$\\
 & (IllustrisTNG) & (IllustrisTNG)  & (SIMBA)  & (SIMBA) & (ASTRID) & (ASTRID)\\
\hline 
$R_{*,1/2}$ vs $M_{*}$ & 79.45 & 0.33  & 127.63  & 0.41 & 581.71 & 3.11\\
$f_{\textrm{DM}}(<R_{*,1/2})$ vs $M_{*}$ & 57.16 & 0.24  & 353.49  & 1.14 & 158.15 & 0.85\\
$M_{\textrm{DM,1/2}}$vs $M_{*}$ & 219.61 & 0.90  & 1270.60  & 4.10 & 558.37 & 2.99\\
$M_{\textrm{tot}}$ vs $M_{*}$ & 212.16 & 0.87  & 170.72  & 0.55 & 290.28 & 1.55\\
cumulative & 568.38 & 2.33  & 1922.45  & 6.20 & 1588.51 & 8.49\\
\hline 
\end{tabular}\label{tab:sim_base_comparison_chi_squared}
\end{table*}

\begin{figure*}
    \centering
\includegraphics[width=\linewidth]{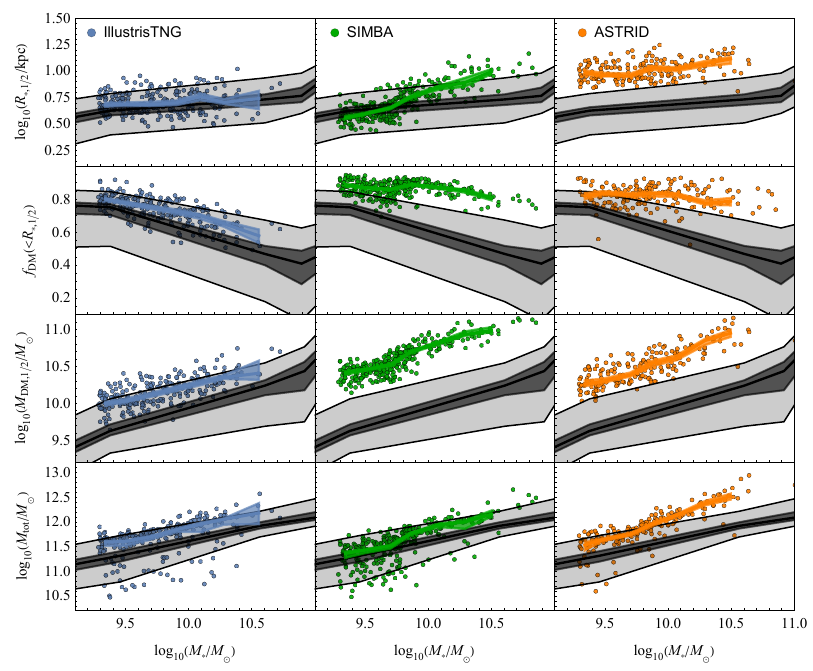}
    \caption{From top to bottom: stellar half-mass radius, $R_{*,1/2}$, DM fraction within $R_{*,1/2}$, $f_{\textrm{DM}}$, DM mass within $R_{*,1/2}$, $M_{\textrm{DM,1/2}}$, and total mass, $M_{\textrm{tot}}$, from \textsc{camels}' fiducial simulations (first column: IllustrisTNG suite, blue points; second column: SIMBA suite, green points; third column: ASTRID suite, orange points), compared with the corresponding SPARC trends, shown as black curves. The shaded grey area represents the scatter of the observed relations, given by the difference between the 16th and the 84th percentiles with the median. The dark grey regions represent the error on the median for the SPARC trends, while the blue, green and orange colored regions represent the error on the median with respect to the IllustrisTNG, SIMBA and ASTRID simulation points, respectively.} \label{fig:sim_base_comparison}
\end{figure*}

Fig. \ref{fig:selection_effects_sims} also shows an important selection effect that could result from analyzing subsequent comparisons between various simulations: as one can see from the right panel, the high-mass threshold between star-forming galaxies (points) and passive galaxies (triangles) is lower for simulations with lower SN feedback values. The difference between the two simulations can initially be accounted for by a straightforward shift along the $M_{*}$ axis, taking into account both star-forming and passive galaxies. However, after applying a cut in specific star-formation rate (points only), this discrepancy can only be explained by an additional shift along the $M_{\textrm{DM}}$ axis. This would erroneously suggest a significant influence of baryonic processes on the dark matter halo properties of the galaxies. It is thus important to note that any apparent effect of the SN feedback processes on DM scaling relations is mainly a combination of effects on the total stellar mass of the galaxies, plus a selection effect due to not considering also the passive galaxies in the trends.

\section{Comparison between observations and \textsc{camels} simulations}\label{sec:Analysis}

The aim of this paper is to compare the scaling relation trends obtained from the SPARC star-forming galaxy sample and the corresponding trends from the \textsc{camels} simulations, in order to determine constraints on cosmological and astrophysical parameters. We start in Section \ref{sec:sim_base_comparison} by comparing the observed trends with the fiducial simulations of all three \textsc{camels} simulation suites. In Section \ref{sec:sim_variable_parameters},  we analyze the effect of varying the cosmological and astrophysical parameters in IllustrisTNG one by one, to check the relative contribution of each parameter independently to the scaling relations. In Sections \ref{sec:CAMELS_best_fit_analysis}-\ref{sec:CAMELS_ASTRID_best_fit_analysis}, to find the combination of cosmological and astrophysical parameters that better fits the observed data, we consider all the 1065 simulations of the IllustrisTNG, SIMBA and ASTRID suites, and perform a chi-squared best fit analysis. We then provide constraints for both cosmological and astrophysical parameters by means of a bootstrapping procedure on both simulations and observations. Finally, in Section \ref{sec:wind_mass_loading_analysis}, we compare the inferred wind mass loading at injection from the IllustrisTNG suite with mass loading trends presented in literature, and compare the trends from the three simulation suites.

\begin{figure}
\includegraphics[width=\columnwidth]{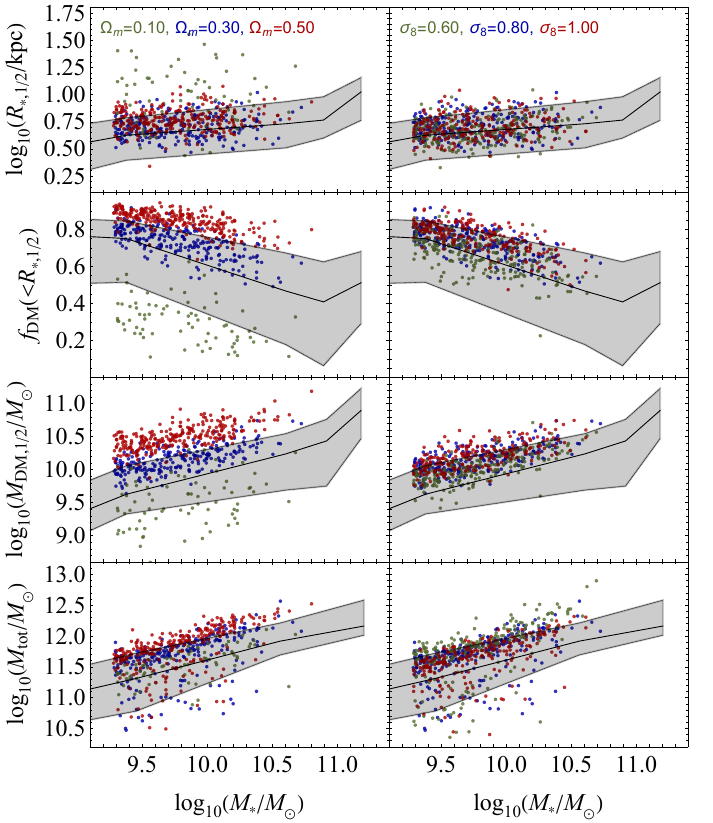}
\caption{Comparison between SPARC observations and IllustrisTNG simulations with differing cosmological parameters. Each row in the plot corresponds to a different scaling relation. Each column shows the effect of varying one of the two cosmological parameters.}
\label{fig:1P_cosmo_variation_comparison}
\end{figure}

\subsection{CAMELS fiducial simulations comparison with SPARC observations}\label{sec:sim_base_comparison}
As a preliminary analysis, it is important to check if the fiducial simulations from IllustrisTNG, SIMBA and ASTRID reproduce accurately the observed trends. We thus compared the simulations labeled `$\textrm{1P\_1\_0}$' from both the IllustrisTNG, SIMBA and ASTRID suites of \textsc{camels}, which are the fiducial simulations, to the observed SPARC scaling relation trends. To obtain the observed trends, we binned the SPARC data in fixed bins of stellar mass, and for each bin we evaluated the 16th, 50th (median) and 84th percentiles. We then linearly interpolated between these points to quantify the observed trends. Discussion on how the binning procedure affects the results is detailed in Appendix \ref{sec:binning_procedure_contribution}.

Fig. \ref{fig:sim_base_comparison} shows the stellar half-mass radius, $R_{*,1/2}$, the DM fraction within the stellar half-mass radius, $f_{\textrm{DM}}(<R_{*,1/2})$, the DM mass within the stellar half-mass radius, $M_{\textrm{DM},1/2}$, and the total mass, $M_{\textrm{tot}}$, as a function of the total stellar mass, $M_{*}$, for star-forming simulated galaxies taken from the IllustrisTNG (blue points), SIMBA (green points) and ASTRID (orange points) suites. The observed SPARC trends for the 16th, median and 84th percentile are shown with black lines, with a shaded grey region representing the scatter of the observed data points. It should be noted that the scatter of the SPARC relation is of the same order of magnitude of the typical observational uncertainties associated to the quantities considered for the scaling relations. The error on the median trends for both SPARC and simulated galaxies' trends are also shown, with a dark grey region for SPARC, and a blue/green/orange region for the IllustrisTNG/SIMBA/ASTRID trends, respectively.

In the IllustrisTNG suite, two of these relations (the size-mass relation and the $M_{\textrm{tot}}$-$M_{*}$ relation) have been used to calibrate the fiducial simulation. Considering the fact that the observed trends used to calibrate these scaling relations in \textsc{camels} do not match exactly the ones we are using in this paper (the SPARC trends), and that the correlations involving the central DM fraction and mass are not used in the calibration of the reference simulations, our results should not be affected by any circularity issue.

All simulations follow the direction of all the trends observed in SPARC, but with an offset with respect to the SPARC trends. This offset qualitatively seems to be stronger for the SIMBA and ASTRID simulations, rather than for the IllustrisTNG simulation. To quantitatively compare the simulations' data to the observed data, we evaluated for the IllustrisTNG, SIMBA and ASTRID simulations and for each of the scaling relations the $\chi^{2}$ between the simulation points and the interpolated SPARC trends, via the following formula:

\begin{equation}
\chi^{2}=\sum_{i=1}^{N_{\textrm{sim}}}\frac{[y_{\textrm{sim, }i}-f_{\textrm{rel}}(x_{\textrm{sim, }i})]^{2}}{\sigma_{\textrm{rel},i}^2},
\end{equation}
\\
where $\chi^{2}$ is the chi-squared evaluated for the scaling relation considered, $(x_{\textrm{sim, }i},\,y_{\textrm{sim, }i})$ are the $N_{\textrm{sim}}$ points from the simulation in the considered scaling relation parameter space, $f_{\textrm{rel}}$ is the observed scaling relation median trend's linear interpolation function, and $\sigma_{\textrm{rel},i}$ is given by the mean between $\sigma_{-}$ and $\sigma_{+}$, which are the differences, in absolute value, between the linear interpolated functions of the 16th and the 84th percentile trends associated to the observed scaling relation, respectively, and the interpolated median trend, each evaluated at $x_{\textrm{sim},i}$,. Given that the various simulations have a different galaxy count $N_{\textrm{sim}}$, to compare different simulations we also considered a normalized chi-squared, defined as $\tilde{\chi}^{2} := \chi^{2}/(N_{\textrm{sim}}-1)$.
The chi-squared for IllustrisTNG, SIMBA and ASTRID fiducial simulations are shown in Table \ref{tab:sim_base_comparison_chi_squared}. 

As seen in Fig. \ref{fig:sim_base_comparison}, comparing the errors on the medians and the $\tilde{\chi}^{2}$ values, only the IllustrisTNG (and SIMBA for log-masses lower than $\sim 5.0\times 10^{9}\,M_{\odot}$) fiducial simulations provide a $R_{*,1/2}$-$M_{\rm *}$ which on average is in agreement with the observations, while the ASTRID simulation is in disagreement with the SPARC trend over the full mass range. All the simulations also produce comparable total masses which are, at fixed stellar mass, slightly larger than the observed values found by \cite{Posti2019}. This discrepancy is stronger for the ASTRID simulation, especially at higher mass values. The size-mass relation results for IllustrisTNG and SIMBA are compatible with the results shown in \cite{Villaescusa2021}, \textit{IX} panel of Fig. 4. The IllustrisTNG fiducial simulation replicates better the scaling relations involving quantities evaluated within the stellar half-mass radius, such as $f_{\textrm{DM}}(<R_{*,1/2})$-$M_{*}$ and $M_{\textrm{DM,1/2}}$-$M_{*}$, with a moderate shift towards higher values at fixed stellar mass. SIMBA and ASTRID fiducial simulations, instead, produce unrealistically high dark matter masses.

\begin{figure*}
    \centering
    \includegraphics[width=0.96\textwidth]{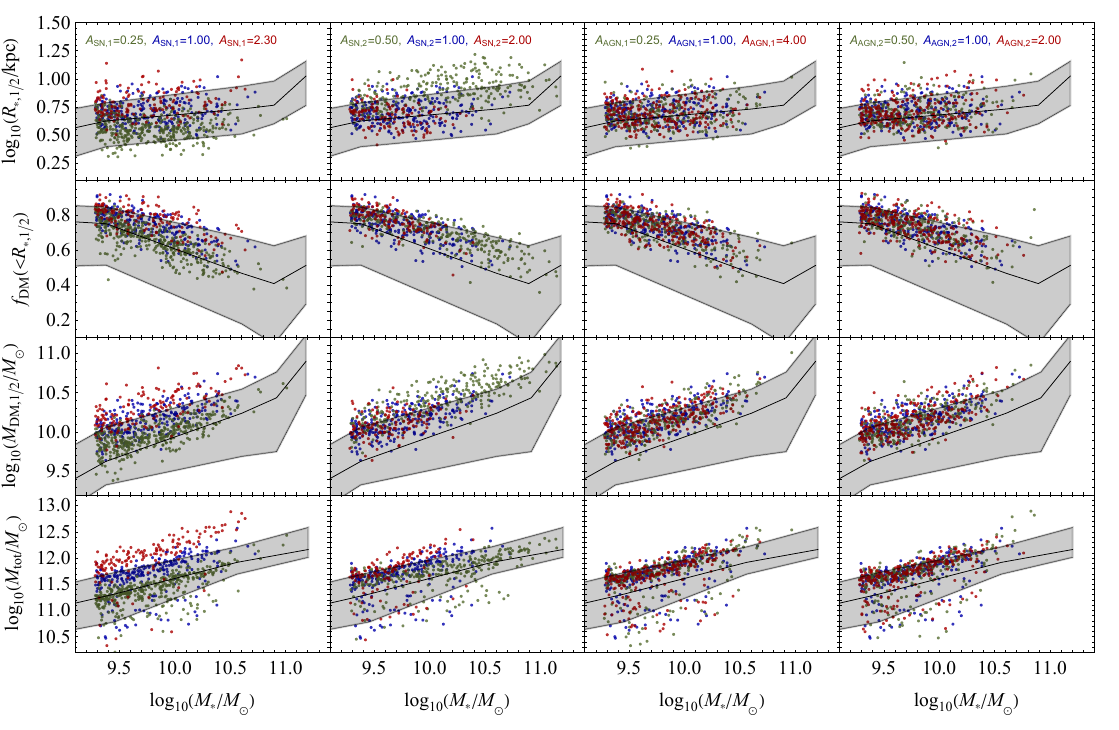}

\caption{Comparison between SPARC observations and IllustrisTNG simulations with differing astrophysical feedback parameters. Each row in the plot corresponds to a different scaling relation. Each column shows the effect of varying one of the four astrophysical parameters.}
    \label{fig:1P_variation_comparison}
\end{figure*}

\subsection{Effects of the variation of cosmological and astrophysical parameters} \label{sec:sim_variable_parameters}

We proceeded by comparing the IllustrisTNG fiducial simulation and the observed SPARC trends with IllustrisTNG simulations from the `1P' simulation set, which assume, for each cosmological and astrophysical parameter, the minimum and maximum value available\footnote{An exception for this has been made for the upper limit of the $A_{\textrm{SN1}}$ parameter. We choose $A_{\textrm{SN1}} = 2.30$ as the upper limit, given the fact that a supernova feedback which is too much energetic suppresses the star formation in almost all galaxies, producing a sample of star-forming galaxies too small to be statistically significant.}.

Fig. \ref{fig:1P_cosmo_variation_comparison} shows the same scaling relations presented in Fig. \ref{fig:sim_base_comparison} for the illustrisTNG simulation, but with each of the columns showing the effects of varying one of the two cosmological parameters on the simulations' trends. Fig. \ref{fig:1P_variation_comparison} is the same as Fig. \ref{fig:1P_cosmo_variation_comparison}, but with each column showing the effects of varying one of the four astrophysical parameters on the simulations' trends, instead. The same figures for SIMBA and ASTRID are shown in Fig. \ref{fig:1P_cosmo_variation_comparison_SIMBA_ASTRID} and \ref{fig:1P_variation_comparison_SIMBA_ASTRID}, and are discussed in detail in Appendix \ref{sec:SIMBA_ASTRID_1P_results}.

Starting with the cosmological parameters, we can see that there is a monotonic trend between increasing values of $\Omega_{\textrm{m}}$ and the normalization of the scaling relations. Moreover, the effects of varying the density parameter, $\Omega_{\textrm{m}}$, on the scaling relation trends are more intense than variations concerning the amplitude of the linear matter density fluctuations, $\sigma_{8}$, for which almost no variation of the scaling relation trends can be appreciated.

Regarding the astrophysical parameters instead, as expected we see that in the range of mass and for the galaxy-type considered, the impact on our scaling relations of the parameters related to the SN feedback is stronger than that of the AGN-related parameters.  For modifying the wind energy per unit star-formation rate, an increase of $A_{\textrm{SN1}}$ corresponds, at fixed stellar mass, to an increase in half-mass radius, dark matter fraction, dark matter mass within the half-mass and total mass. As we have already shown in Fig. \ref{fig:selection_effects_sims}, $A_{\textrm{SN1}}$ impacts strongly the stellar mass accretion, which would explain most of the changes observed \footnote{This is confirmed by the dependence of the star formation density as a function of redshift and astrophysical parameters in Figure 9 of \citealt{Villaescusa2021}.}. Of course, more energetic winds are expected to push the gas to larger distances, altering the gravitational potential, the half-mass radii, and thus making DM mass and DM fraction larger. Therefore, with all the other parameters fixed to the reference values, less energetic models better reproduce the observations.

The effects of increasing the wind speed at injection ($A_{\textrm{SN2}}$) are instead more subtle. While the internal DM fractions and total mass are practically unchanged, with only a slight slope change at $M_{*} \sim 1.6\times 10^{10}\,\textrm{M}_{\odot}$ for the $M_{\textrm{tot}}$-$M_{*}$ relation, there is an increase in stellar half-mass radius for the simulation with lower wind speeds at injection. This increase seems to be stronger for star-forming galaxies of intermediate-high mass. The increase in stellar half-mass radius also implies an increased DM mass within the stellar half-mass radius.  For low values of $A_{\textrm{SN2}}$, the scaling relations extend to much higher values of stellar mass, since winds with less momentum allow the formation of very massive star-forming galaxies. On the other hand, higher values of $A_{\textrm{SN2}}$ quench more efficiently star formation, preventing the formation of more massive galaxies, which results in scaling relation trends that stop at lower stellar mass. These trends are only mildly seen varying $A_{\textrm{SN1}}$. 

The effects of changing $A_{\textrm{AGN1}}$ and $A_{\textrm{AGN2}}$ instead seem to be negligible: we cannot notice any apparent change of normalization, slope or scatter in the scaling relation trends among the two extreme values adopted for the two parameters.

In regards to SIMBA and ASTRID results, for the cosmological parameters we find that in both cases an increase in $\Omega_{\textrm{m}}$ corresponds to an increase in the normalization of the scaling relations. In both cases, there is better concordance with the observations for low values of $\Omega_{\textrm{m}}$, but also a reduction in the number of late-type galaxies (LTGs) present in the simulations. Neither suite instead shows sensitivity to variations of $\sigma_{8}$. For the astrophysical parameters, the analysis done in Appendix \ref{sec:SIMBA_ASTRID_1P_results} shows that in both cases no simulation which is only subject to the variation of one astrophysical parameter can reconcile the simulated galaxies' trends with the observed SPARC trends, especially for scaling relations relative to central DM masses and DM fractions. In the SIMBA case, one necessarily needs to lower $\Omega_{\textrm{m}}$, while in ASTRID low values of $\hat{A}_{\textrm{SN2}}$ solves the discrepancy, but at the cost of having all galaxies clustered at low stellar mass values.

\subsection{IllustrisTNG simulations' best-fit to the observations}\label{sec:CAMELS_best_fit_analysis}

From the analyses performed in the previous sections, it can be seen that the fiducial simulations do not exactly reproduce the observed trends, and that varying the astrophysical and cosmological parameters could improve the agreement. Therefore, we searched within the IllustrisTNG suite simulations for the set of cosmological and astrophysical parameters that provide a best-fit to the observed SPARC trends. We considered all the 1065 simulations from the `LH', `1P' and `EX' sets, and for each of the simulations we followed the same procedure detailed in Section \ref{sec:sim_base_comparison}. We then ordered the simulations according to the value of the respective cumulative $\tilde{\chi}^{2}$ result.

\begin{figure*}
    \centering
    \includegraphics[width=\linewidth]{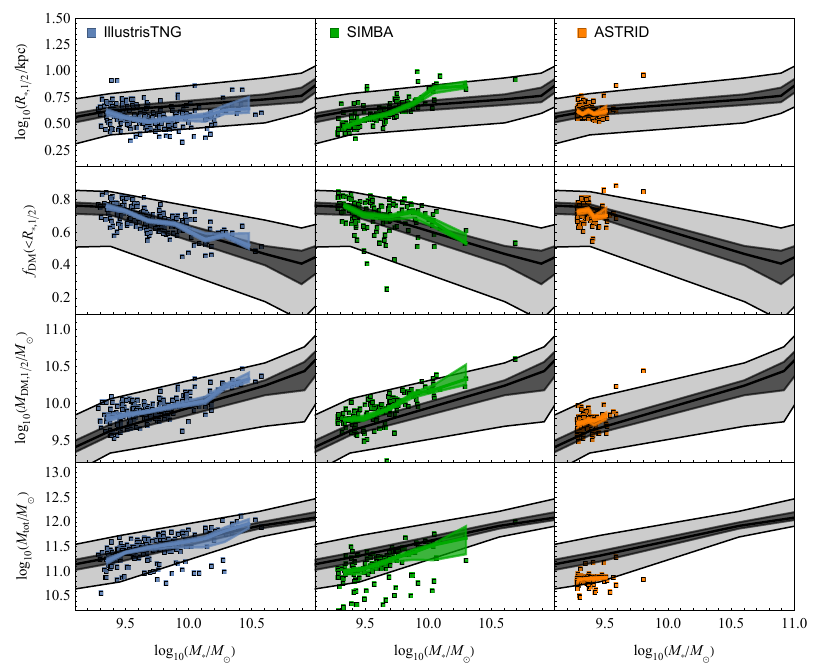}
    \caption{Same as Fig. \ref{fig:sim_base_comparison}, but comparing the \textsc{camels} best-fit simulations.}
    \label{fig:best_fit_sim_comparison}
\end{figure*}

We find that the simulation that better fits all the observed SPARC data is the simulation `LH-698', having the following cosmological and astrophysical parameters: $\Omega_{\textrm{m}} = 0.27$, $\sigma_{8} = 0.83$, $S_{8} = 0.78$, $A_{\textrm{SN1}} = 0.48$, $A_{\textrm{SN2}} = 1.24$, $A_{\textrm{AGN1}} = 2.53$ and $A_{\textrm{AGN2}} = 1.79$, where the value of $S_{8}$ has been inferred from $\Omega_{\textrm{m}}$ and $\sigma_{8}$  via the definition, $S_{8} := \sigma_{8} \sqrt{\Omega_{\textrm{m}}/0.3}$. The normalized chi-squared associated to this simulation is $\tilde{\chi}^{2} = 1.17$. The first column of Fig. \ref{fig:best_fit_sim_comparison} shows the comparison between this simulation and the observed SPARC trends. The fact that the best-fit simulation obtained is not one of the fiducial simulations seems to reassure against eventual circularity problems in this procedure.

It has to be noted that there is the chance that other simulations in the IllustrisTNG suite, with different parameter combinations, show a similar chi-squared as the one of the `best-fit' simulation considered above. This is because different parameter combinations, by compensation with each other due to degeneracies, could give rise to similar physical conditions for the galaxies, and thus produce similar scaling relations with respect to the ones observed in our Universe. Indeed, by employing a method of Bayesian inference based on implicit likelihood inference (ILI), by using the observed star formation rate density (SFRD) and, separately, the stellar mass functions (SMFs), at different redshifts, \cite{Yongseok2023} confirm the existence of degeneracies between cosmological and astrophysical parameters in \textsc{camels}.

To check that the choice of parameters associated with the best-fit simulation is not just the result of a statistical fluctuation, and to assign statistical uncertainties to the parameters, we decided to perform a bootstrap analysis of the best-fit sample, which enabled us to take into account the uncertainty induced by the degeneracies among fitted parameters.

To verify that the procedure recovers the ground-truth correctly within a certain confidence limit, we have first tested it by using mock observational data taken from the `1P\_1\_0' fiducial simulation and various LH simulations, instead of real data. The results confirm that this procedure performs well, recovering the ground truth in all of the cases tested. Test results show that the parameters that are better constrained by this approach are $\Omega_{\textrm{m}}$ and $A_{\textrm{SN1}}$, while the constraining power for $\sigma_{8}$, $S_{8}$ and $A_{\textrm{SN2}}$ is milder. AGN-feedback related parameters are instead roughly constrained by this method. These results confirm the dependencies found in Section \ref{sec:sim_variable_parameters}. More details are provided in Appendix \ref{sec:consistency test}.

We subsequently applied this method with the SPARC catalog as the observational data, by bootstrapping both the simulations and the observed data, with the aim of obtaining constraints on both the cosmological and the astrophysical parameters from the sample of best-fit simulations, and not from just one simulation. We bootstrapped each of the 1065 simulations and the SPARC dataset 100 times\footnote{The bootstrapping process is performed via the Mathematica resource function "BootstrapStatistics": \url{https://resources.wolframcloud.com/FunctionRepository/resources/BootstrapStatistics/}. The mean fraction of substitutions with duplicate elements over the total number of objects in the bootstrapped array that this function performs is constant, and equal to $\simeq 0.36$.}, and for each of the resamplings we performed the same analysis detailed in Section \ref{sec:sim_base_comparison}. We then order, for each resampling, the simulations according to the values of $\tilde{\chi}^{2}$, and take the best-fit simulation. We thus obtained a list of 100 best-fit simulations. The constraints obtained, associated to each of the correlations, are summarized in Table \ref{tab:CAMELS_bootstrap_analysis_results}. The constraints are given in terms of the 16th, 50th (median) and 84th percentiles.

\begin{table*}
\centering
\renewcommand*{\arraystretch}{1.50}
\caption{Constraints on cosmological and astrophyiscal parameters, based on the methods detailed in Section \ref{sec:CAMELS_best_fit_analysis}, for the IllustrisTNG suite. The constraints are given in terms of 16th, 50th and 84th percentiles. The `cumulative alt.' column values are the cumulative chi-squared results obtained by considering only the internal dark matter scaling relations.}
\begin{tabular}{ccccccc}
\hline
\hline
Parameter & $R_{*,1/2}$-$M_{*}$ & $f_{\textrm{DM}}(<R_{*,1/2})$-$M_{*}$ & $M_{\textrm{DM},1/2}$-$M_{*}$ & $M_{\textrm{tot}}$-$M_{*}$  & cumulative & cumulative alt.\\
\hline 
$\Omega_{\textrm{m}}$ & $0.29_{-0.04}^{+0.05}$ & $0.24_{-0.04}^{+0.03}$ & $0.21_{-0.02}^{+0.02}$  & $0.36_{-0.05}^{+0.10}$  & $0.27_{-0.05}^{+0.01}$ & $0.22_{-0.02}^{+0.03}$\\
$\sigma_{8}$ & $0.85_{-0.16}^{+0.11}$ & $0.83_{-0.03}^{+0.11}$ & $0.87_{-0.07}^{+0.10}$  & $0.95_{-0.04}^{+0.03}$ & $0.83_{-0.11}^{+0.08}$ & $0.92_{-0.10}^{+0.05}$\\
$S_{8}$ & $0.82_{-0.16}^{+0.14}$ & $0.78_{-0.07}^{+0.05}$ & $0.74_{-0.06}^{+0.06}$ & $1.05_{-0.20}^{+0.16}$ & $0.78_{-0.09}^{+0.03}$ & $0.79_{-0.07}^{+0.01}$\\
$A_{\textrm{SN1}}$ & $0.70_{-0.32}^{+0.44}$ & $0.48_{-0.12}^{+0.73}$ & $0.32_{-0.01}^{+0.10}$  & $0.35_{-0.07}^{+0.07}$ & $0.48_{-0.16}^{+0.25}$ & $0.37_{-0.09}^{+0.11}$\\
$A_{\textrm{SN2}}$ & $1.17_{-0.23}^{+0.27}$ & $1.24_{-0.38}^{+0.44}$ & $0.76_{-0.25}^{+0.47}$  & $1.00_{-0.32}^{+0.10}$  & $1.21_{-0.34}^{+0.03}$ & $0.80_{-0.22}^{+0.60}$\\
$A_{\textrm{AGN1}}$ & $2.09_{-1.09}^{+1.46}$ & $1.41_{-0.93}^{+1.12}$ & $1.35_{-0.64}^{+0.62}$  & $1.26_{-0.95}^{+2.30}$  & $2.53_{-1.82}^{+0.89}$ & $1.36_{-0.90}^{+1.17}$\\
$A_{\textrm{AGN2}}$ & $1.39_{-0.63}^{+0.60}$ & $1.46_{-0.64}^{+0.33}$ & $1.25_{-0.68}^{+0.38}$  & $0.83_{-0.07}^{+0.47}$  & $1.31_{-0.67}^{+0.49}$ & $1.37_{-0.71}^{+0.43}$\\
$\chi^{2}$ & $60_{-26}^{+41}$ & $22_{-09}^{+17}$ & $45_{-16}^{+16}$ & $89_{-40}^{+46}$  & $344_{-112}^{+185}$ & $99_{-28}^{+33}$\\
$\tilde{\chi}^{2}$ & $0.25_{-0.06}^{+0.06}$ & $0.10_{-0.03}^{+0.03}$ & $0.14_{-0.03}^{+0.03}$  & $0.28_{-0.10}^{+0.12}$  & $1.23_{-0.20}^{+0.29}$ & $0.32_{-0.07}^{+0.08}$\\
\hline
\end{tabular}
\label{tab:CAMELS_bootstrap_analysis_results}
\end{table*}

We obtain $\Omega_{\textrm{m}} = 0.27_{-0.05}^{+0.01}$, $\sigma_{8} = 0.83_{-0.11}^{+0.08}$, $S_{8} = 0.78_{-0.09}^{+0.03}$, $A_{\textrm{SN1}} = 0.48_{-0.16}^{+0.25}$, $A_{\textrm{SN2}} = 1.21_{-0.34}^{+0.03}$, $A_{\textrm{AGN1}} = 2.53_{-1.82}^{+0.89}$ and $A_{\textrm{AGN2}} = 1.31_{-0.67}^{+0.49}$, with an associated normalized chi-squared of $\tilde{\chi}^{2} = 1.23_{-0.20}^{+0.29}$.

While we manage to constrain the cosmological and SN feedback parameters, we are unable to constrain the AGN feedback parameters. This latter result is expected and consistent with the trends discussed in Sec. \ref{sec:sim_variable_parameters}. Regarding the cosmological parameters, $\Omega_{\textrm{m}}$ and $S_{8}$ are better constrained than $\sigma_{8}$, while in the case of the SN feedback parameters, $A_{\textrm{SN1}}$ is better constrained than $A_{\textrm{SN2}}$.

To further analyze the impact of eventual circularity effects on our results, we also report in Table \ref{tab:CAMELS_bootstrap_analysis_results} the cumulative chi-squared results obtained by considering only the internal dark matter scaling relation, that is, the $f_{\textrm{DM}}$-$M_{*}$ and the $M_{\textrm{DM,1/2}}$-$M_{*}$ relations. In this case, the results are $\Omega_{\textrm{m}} = 0.22_{-0.02}^{+0.03}$, $\sigma_{8} = 0.92_{-0.10}^{+0.05}$, $S_{8} = 0.79_{-0.07}^{+0.01}$, $A_{\textrm{SN1}} = 0.37_{-0.09}^{+0.11}$, $A_{\textrm{SN2}} = 0.80_{-0.22}^{+0.60}$, $A_{\textrm{AGN1}} = 1.36_{-0.90}^{+1.17}$ and $A_{\textrm{AGN2}} = 1.37_{-0.71}^{+0.43}$, which are compatible with the cumulative results within $1\sigma$.

\subsection{SIMBA simulations' best-fit to the observations}\label{sec:CAMELS_SIMBA_best_fit_analysis}
We considered all the 1065 `LH', `1P' and `EX' simulations from the SIMBA suite, to check if there is a simulation with a set of reasonable cosmological and astrophysical parameters that fits the observations. By repeating the same procedure detailed in Section \ref{sec:CAMELS_best_fit_analysis}, we found that the best-fit SIMBA simulation for all the SPARC observed trends is the simulation `LH-360', having the following parameters: $\Omega_{\textrm{m}} = 0.13$, $\sigma_{8} = 1.00$, $S_{8} = 0.65$, $\tilde{A}_{\textrm{SN1}} = 0.35$, $\tilde{A}_{\textrm{SN2}} = 0.50$, $\tilde{A}_{\textrm{AGN1}} = 0.68$ and $\tilde{A}_{\textrm{AGN2}} = 1.16$, with a normalized chi-squared of $\tilde{\chi}^{2} = 1.98$. The second column of Fig. \ref{fig:best_fit_sim_comparison} shows the comparison between this simulation and the observed SPARC trends.

We performed again the bootstrap analysis detailed in Section \ref{sec:CAMELS_best_fit_analysis}, this time on both the SIMBA simulations and the SPARC dataset. Results are summarized in Table \ref{tab:CAMELS_SIMBA_bootstrap_analysis_results}.

\begin{table*}
\centering
\renewcommand*{\arraystretch}{1.50}
\caption{Same as Table \ref{tab:CAMELS_bootstrap_analysis_results}, but for the SIMBA suite. The constraints are given in terms of 16th, 50th and 84th percentiles.}
\begin{tabular}{cccccc}
\hline
\hline
Parameter & $R_{*,1/2}$-$M_{*}$ & $f_{\textrm{DM}}(<R_{*,1/2})$-$M_{*}$ & $M_{\textrm{DM},1/2}$-$M_{*}$  & $M_{\textrm{tot}}$-$M_{*}$  & cumulative\\
\hline 
$\Omega_{\textrm{m}}$ & $0.45_{-0.25}^{+0.01}$ & $0.14_{-0.01}^{+0.01}$ & $0.11_{-0.01}^{+0.01}$  & $0.46_{-0.05}^{+0.03}$  & $0.14_{-0.01}^{+0.02}$\\
$\sigma_{8}$ & $0.961_{-0.064}^{+0.002}$ & $0.98_{-0.23}^{+0.02}$ & $0.76_{-0.01}^{+0.24}$  & $0.96_{-0.06}^{+0.02}$  & $0.96_{-0.25}^{+0.03}$\\
$S_{8}$ & $1.18_{-0.445}^{+0.004}$ & $0.65_{-0.14}^{+0.00}$ & $0.49_{-0.03}^{+0.17}$ & $1.18_{-0.06}^{+0.06}$ & $0.649_{-0.135}^{+0.004}$\\
$\tilde{A}_{\textrm{SN1}}$ & $2.47_{-2.12}^{+0.04}$ & $0.40_{-0.05}^{+0.18}$ & $0.44_{-0.09}^{+0.20}$  & $1.49_{-0.98}^{+1.02}$  & $0.45_{-0.10}^{+0.06}$\\
$\tilde{A}_{\textrm{SN2}}$ & $1.63_{-0.95}^{+0.21}$ & $0.78_{-0.28}^{+1.06}$ & $0.71_{-0.21}^{+0.83}$  & $0.69_{-0.17}^{+0.43}$  & $0.81_{-0.31}^{+1.03}$\\
$\tilde{A}_{\textrm{AGN1}}$ & $0.77_{-0.50}^{+0.16}$ & $0.68_{-0.14}^{+0.00}$ & $0.68_{-0.20}^{+0.55}$  & $0.65_{-0.26}^{+0.87}$  & $0.56_{-0.02}^{+0.12}$\\
$\tilde{A}_{\textrm{AGN2}}$ & $1.28_{-0.71}^{+0.32}$ & $1.16_{-0.04}^{+0.01}$ & $1.16_{-0.20}^{+0.10}$  & $1.30_{-0.59}^{+0.35}$  & $1.13_{-0.35}^{+0.03}$\\
$\chi^{2}$ & $34_{-13}^{+20}$ & $27_{-12}^{+13}$ & $22_{-06}^{+12}$  & $63_{-21}^{+33}$  & $261_{-083}^{+109}$\\
$\tilde{\chi}^{2}$ & $0.14_{-0.04}^{+0.06}$ & $0.21_{-0.07}^{+0.10}$ & $0.23_{-0.08}^{+0.06}$  & $0.19_{-0.06}^{+0.07}$  & $2.01_{-0.43}^{+0.52}$\\
\hline
\end{tabular}
\label{tab:CAMELS_SIMBA_bootstrap_analysis_results}
\end{table*}

We obtain $\Omega_{\textrm{m}} = 0.14_{-0.01}^{+0.02}$, $\sigma_{8} = 0.96_{-0.25}^{+0.03}$, $S_{8} = 0.649_{-0.135}^{+0.004}$, $\tilde{A}_{\textrm{SN1}} = 0.45_{-0.10}^{+0.06}$, $\tilde{A}_{\textrm{SN2}} = 0.81_{-0.31}^{+1.03}$, $\tilde{A}_{\textrm{AGN1}} = 0.56_{-0.02}^{+0.12}$ and $\tilde{A}_{\textrm{AGN2}} = 1.13_{-0.35}^{+0.03}$, with an associated normalized chi-squared of $\tilde{\chi}^{2} = 2.01_{-0.43}^{+0.52}$.

In the case of the SIMBA suite, we are unable to give meaningful constraints on the SN feedback parameter $\tilde{A}_{\textrm{SN2}}$, but we manage to constrain the two AGN feedback parameters $\tilde{A}_{\textrm{AGN1}}$ and $\tilde{A}_{\textrm{AGN2}}$, which in this case are associated to the physical properties of the SMBH jets. Once again, the cosmological parameters that are better constrained are $\Omega_{\textrm{m}}$ and $S_{8}$, while $\sigma_{8}$ has a higher associated uncertainty. However, these results are obtained at the cost of considering values of $\Omega_{\textrm{m}}$ that are near 0.10 and of $\sigma_{8}$ near 1.00. We have verified that, by lowering the value of $\Omega_{\textrm{m}}$, the dependence of the scaling relations from the SN and AGN feedback parameters are different with respect to what was shown in Figure \ref{fig:1P_variation_comparison_SIMBA_ASTRID}, which was evaluated for the reference cosmology. In particular, the dependence on the AGN feedback parameters is stronger. A possible motivation could be that, for a lower value of $\Omega_{\textrm{m}}$, the halos are less massive and thus the momentum transfer between the SMBH jets and the host galaxy particles is more effective in SIMBA compared to IllustrisTNG, in which the isotropic kinetic feedback results in isotropic winds with lower velocities. Similarly to the results from Table \ref{tab:sim_base_comparison_chi_squared}, we again find that SIMBA shows, on average, a worse agreement with the observed values than IllustrisTNG, having a higher cumulative $\tilde{\chi}^{2}$ value.

\subsection{ASTRID simulations' best-fit to the observations}\label{sec:CAMELS_ASTRID_best_fit_analysis}
Finally, we considered all the 1061 `LH' and `1P' simulations of the ASTRID suite, and performed the same analysis as in Sections \ref{sec:CAMELS_best_fit_analysis} and \ref{sec:CAMELS_SIMBA_best_fit_analysis}. We found that the best-fit ASTRID simulation for all the SPARC observed trends is the simulation `LH-474', having the following parameters: $\Omega_{\textrm{m}} = 0.46$, $\sigma_{8} = 0.97$, $S_8 = 1.20$, $\hat{A}_{\textrm{SN1}} = 0.42$, $\hat{A}_{\textrm{SN2}} = 0.59$, $\hat{A}_{\textrm{AGN1}} = 1.58$ and $\hat{A}_{\textrm{AGN2}} = 0.64$, with a normalized chi-squared of $\tilde{\chi}^{2} = 1.52$. The third column of Fig. \ref{fig:best_fit_sim_comparison} shows the comparison between this simulation and the observed SPARC trends.

We also performed the bootstrap analysis detailed in Section \ref{sec:CAMELS_best_fit_analysis}, on both ASTRID simulations and the SPARC dataset. Results are summarized in Table \ref{tab:CAMELS_ASTRID_bootstrap_analysis_results}.

\begin{table*}
\centering
\renewcommand*{\arraystretch}{1.50}
\caption{Same as Table \ref{tab:CAMELS_bootstrap_analysis_results}, but for the ASTRID suite. The constraints are given in terms of 16th, 50th and 84th percentiles.}
\begin{tabular}{cccccc}
\hline
\hline
Parameter & $R_{*,1/2}$-$M_{*}$ & $f_{\textrm{DM}}(<R_{*,1/2})$-$M_{*}$ & $M_{\textrm{DM},1/2}$-$M_{*}$  & $M_{\textrm{tot}}$-$M_{*}$  & cumulative\\
\hline 
$\Omega_{\textrm{m}}$ & $0.41_{-0.09}^{+0.03}$ & $0.27_{-0.14}^{+0.22}$ & $0.44_{-0.26}^{+0.00}$ & $0.13_{-0.01}^{+0.24}$ & $0.44_{-0.15}^{+0.02}$ \\
$\sigma_{8}$ & $0.87_{-0.22}^{+0.05}$ & $0.85_{-0.18}^{+0.12}$ & $0.64_{-0.00}^{+0.25}$ & $0.87_{-0.20}^{+0.06}$ & $0.81_{-0.17}^{+0.15}$\\
$S_{8}$ & $0.98_{-0.20}^{+0.03}$ & $0.80_{-0.30}^{+0.37}$ & $0.78_{-0.23}^{+0.11}$ & $0.57_{-0.13}^{+0.29}$ & $0.85_{-0.06}^{+0.36}$\\
$\hat{A}_{\textrm{SN1}}$ & $0.38_{-0.13}^{+0.07}$ & $0.89_{-0.52}^{+2.39}$ & $0.25_{-0.00}^{+0.66}$ & $1.59_{-0.91}^{+0.39}$ & $0.41_{-0.17}^{+0.34}$\\
$\hat{A}_{\textrm{SN2}}$ & $0.55_{-0.04}^{+0.18}$ & $0.63_{-0.06}^{+1.06}$ & $0.73_{-0.19}^{+0.00}$ & $0.64_{-0.03}^{+0.19}$ & $0.61_{-0.04}^{+0.12}$\\
$\hat{A}_{\textrm{AGN1}}$ & $1.58_{-0.79}^{+0.91}$ & $0.82_{-0.47}^{+1.14}$ & $2.48_{-1.61}^{+0.00}$ & $0.73_{-0.09}^{+2.90}$ & $2.49_{-0.90}^{+0.56}$\\
$\hat{A}_{\textrm{AGN2}}$ & $0.58_{-0.05}^{+0.09}$ & $0.81_{-0.25}^{+0.48}$ & $0.52_{-0.00}^{+0.21}$ & $0.56_{-0.03}^{+0.20}$ & $0.62_{-0.09}^{+1.05}$\\
$\chi^{2}$ & $11_{-10}^{+16}$ & $3_{-4}^{+24}$ & $2_{-1}^{+10}$ & $1_{-1}^{+2}$ & $115_{-67}^{+442}$\\
$\tilde{\chi}^{2}$ & $0.07_{-0.03}^{+0.03}$ & $0.07_{-0.04}^{+0.06}$ & $0.06_{-0.02}^{+0.02}$ & $0.06_{-0.03}^{+0.05}$ & $1.65_{-0.40}^{+0.58}$\\
\hline
\end{tabular}
\label{tab:CAMELS_ASTRID_bootstrap_analysis_results}
\end{table*}

We obtain $\Omega_{\textrm{m}} = 0.44_{-0.15}^{+0.02}$, $\sigma_{8} = 0.81_{-0.17}^{+0.15}$, $S_{8} = 0.85_{-0.06}^{+0.36}$, $\hat{A}_{\textrm{SN1}} = 0.41_{-0.17}^{+0.34}$, $\hat{A}_{\textrm{SN2}} = 0.61_{-0.04}^{+0.12}$, $\hat{A}_{\textrm{AGN1}} = 2.49_{-0.90}^{+0.56}$ and $\hat{A}_{\textrm{AGN2}} = 0.62_{-0.09}^{+1.05}$, with a normalized chi-squared of $\tilde{\chi}^{2} = 1.65_{-0.40}^{+0.58}$.

In the case of the ASTRID suite, we have that both $\Omega_{\textrm{m}}$ and $\sigma_{8}$ have large uncertainties, with the upper uncertainty on $\Omega_{\textrm{m}}$ lower than the one of $\sigma_{8}$. The value of $\Omega_{\textrm{m}}$ is very high with respect to the values found in both IllustrisTNG and SIMBA results, while the value of $\sigma_{8}$ is compatible with both \cite{Planck2018} and \cite{Hinshaw2013} results.

As far as the astrophysical parameters are concerned, both SN-feedback parameters are significantly lower than the fiducial value, with $\hat{A}_{\textrm{SN2}}$ showing lower uncertainty than $\hat{A}_{\textrm{SN1}}$. We also find that the parameter $\hat{A}_{\textrm{AGN1}}$ is poorly constrained, while we cannot constrain the parameter $\hat{A}_{\textrm{AGN2}}$.

As shown in Fig. \ref{fig:best_fit_sim_comparison}, these results are obtained at the cost of having a value of $\Omega_{\textrm{m}}$ close to $0.40$ and all galaxies confined in a small region around $M_{*}\sim2\times10^{9}\,M_{\odot}$. The latter is a very similar behavior to the simulation with $\hat{A}_{\textrm{SN2}} = 0.50$, shown in Appendix \ref{sec:SIMBA_ASTRID_1P_results}, which could imply that it is an effect associated to the fact that, differently from equation \eqref{eq:IllustrisTNG_wind_velocity}, the wind velocity in ASTRID does not have a wind velocity floor, thus allowing very low values of $v_{\textrm{w}}$, or could be an effect that depends on other parameters, for example a low value of $\hat{A}_{\textrm{AGN2}}$ (or a mix of these causes). A speculative mechanism that tries to explain why only low-mass LTGs remain in these simulations is presented in Section \ref{sec:1P_results_interpretations}.

\subsection{Wind mass loading analysis}\label{sec:wind_mass_loading_analysis}

Given that the wind mass loading factor is one of the principal quantities that is influenced by the SN feedback parameters and enters in any chemical evolution model (e.g., \citealt{Peeples2011,Tortora2022}), it is important to check its trends and compare the results from both suites and with literature results.

\begin{figure*}
\centering

    \includegraphics[width=\linewidth]{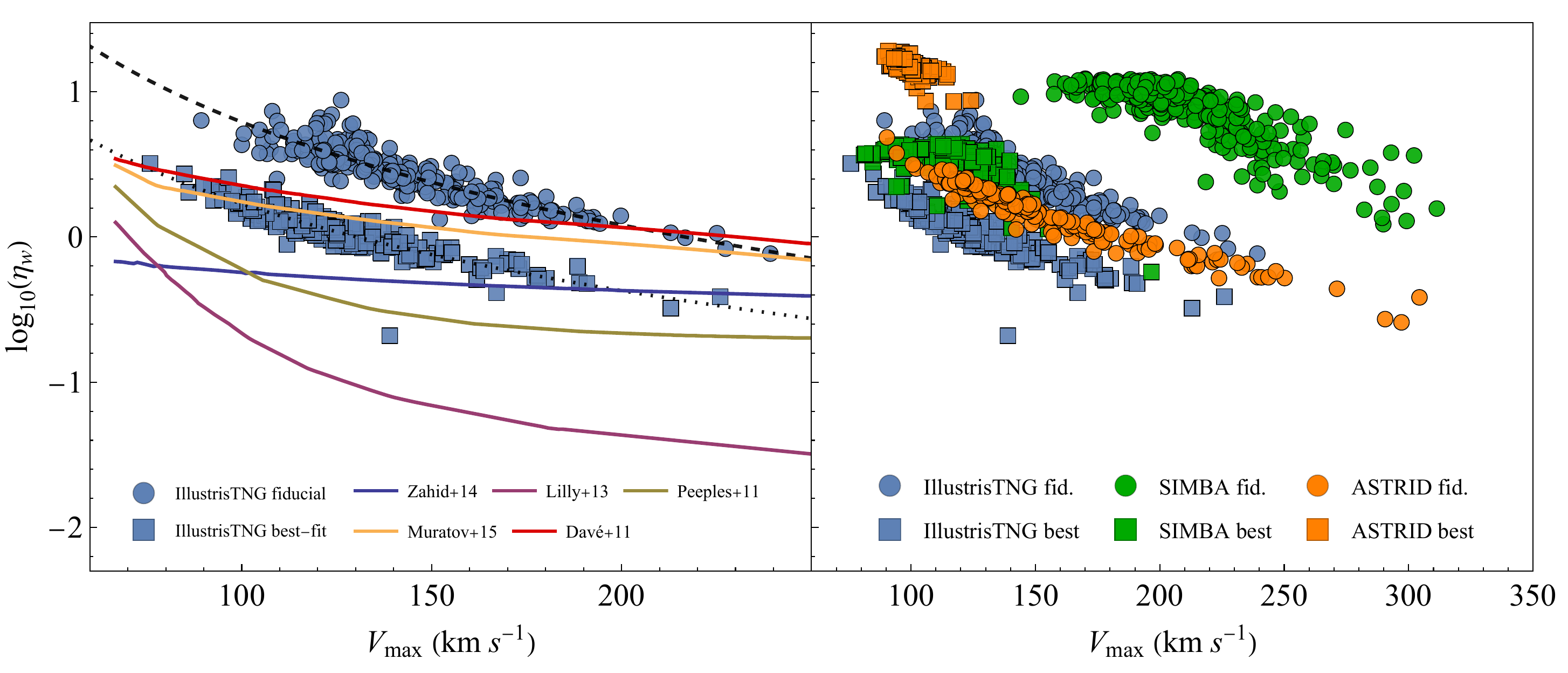}
    
\caption{\textit{Left panel}. Wind mass loading at injection from IllustrisTNG fiducial simulation (blue points) and from the best-fit `$\textrm{LH-698}$' simulation (blue squares), compared with mass loading factors measured in the hydrodynamical simulations described in \protect\cite{Davé2011} and \protect\cite{Muratov2015} (red and orange lines) and extracted from \protect\cite{Belfiore2019}. We also show the empirical determinations of the mass loading factor inferred from measurements of the local mass–metallicity relation by \protect\cite{Peeples2011} (ocher line), \protect\cite{Lilly2013} (dark violet line) and \protect\cite{Zahid2014} (dark blue line). Dashed and dotted black lines are the best fit trends for the fiducial and best-fit IllustrisTNG simulations, respectively. \textit{Right panel}. Comparison between the wind mass loading at injection trends obtained from the IllustrisTNG fiducial simulation (blue points), the IllustrisTNG best-fit simulation (blue squares), the SIMBA fiducial simulation (green points), the SIMBA best-fit simulation (green squares), the ASTRID fiducial simulation (orange points) and the ASTRID best-fit simulation (orange squares).}
\label{fig:wind_mass_loading_comparison}
\end{figure*}

In the left panel of Fig.  \ref{fig:wind_mass_loading_comparison} we show the mass loading factor at injection from the IllustrisTNG suite, taken by evaluating Eqs. \eqref{eq:IllustrisTNG_energy_unit_SFR} and \eqref{eq:IllustrisTNG_wind_velocity} numerically for each galaxy by using $\overline{\sigma}_{\textrm{DM}}$ and $Z$, as a function of the maximum velocity of the rotation curve, $V_{\textrm{max}}$. These trends are obtained by considering both the fiducial simulation and the best-fit simulation as determined in Section \ref{sec:CAMELS_best_fit_analysis}. It emerges that the IllustrisTNG best fit trend is, on average, $0.60\;\textrm{dex}$ lower than the fiducial counterpart, with respect to $\eta_{\textrm{w}}$ values. As discussed in Sec. \ref{sec:sim_variable_parameters}, this discrepancy can be explained by the fact that, in the best-fit simulations, galactic wind outflows are overall less energetic.

We find that the fiducial simulation is compatible with trends measured in hydrodynamical simulations described in \cite{Davé2011} and \cite{Muratov2015} at high values of $V_{\textrm{max}}$, but is totally incompatible with the empirical determinations of the mass loading factor, inferred from measurements of the mass–metallicity relation, presented in \cite{Peeples2011}, \cite{Lilly2013} or \cite{Zahid2014}, while the best-fit `$\textrm{LH-698}$' simulation is placed in between the trends of \cite{Muratov2015} and \cite{Zahid2014}, and is compatible with the former at low values of $V_{\textrm{max}}$.

The right panel of Fig. \ref{fig:wind_mass_loading_comparison} shows instead the comparison between the three simulation suites' wind mass loading trends as a function of maximum rotational velocity. The trends from SIMBA have been obtained by plotting the $\eta_{\textrm{w}}$, evaluated for each galaxy by considering the respective stellar mass values, against the associated $V_{\textrm{max}}$ values, while the trends from ASTRID have been obtained in a manner similar to IllustrisTNG, but using equations \eqref{eq:ASTRID_wind_mass_loading_trend} and \eqref{eq:ASTRID_wind_velocity} instead. For evaluating the ASTRID points, we also had to use the cumulative velocity dispersion, $\sigma$, instead of $\overline{\sigma}_{\textrm{DM}}$, because the values of the one-dimensional local dark matter velocity dispersion for each star particle are not provided for ASTRID in the \textsc{camels} suite. We have checked with direct comparisons in IllustrisTNG that the difference between using $\sigma$ or $\overline{\sigma}_{\textrm{DM}}$ on the mass loading values amounts to an overestimate of the mass loading values of no more than $0.2\,\textrm{dex}$ when using $\sigma$, compared to using $\overline{\sigma}_{\textrm{DM}}$.

As one can see, the fiducial SIMBA mass loading values are, on average, $0.51\;\textrm{dex}$ higher and shifted towards higher velocities than the IllustrisTNG fiducial trend. The best-fit trend tends to agree better with the IllustrisTNG simulations, but (as we saw in Section \ref{sec:CAMELS_SIMBA_best_fit_analysis}) this is achieved by using unreasonable values of the cosmological parameters, along with a lower wind mass loading factor parameter, $\tilde{A}_{\textrm{SN1}}$. The very low value of $\Omega_\textrm{m}$ in the best-fit simulation is strongly impacting the formation of very massive halos, preventing their formation, contrary to what happens in the reference SIMBA simulation.

It should be noted that, as discussed in Section \ref{sec:CAMELS_simulations}, the discrepancy between \cite{Muratov2015}'s mass loading trend (orange curve in the left panel of Fig. \ref{fig:wind_mass_loading_comparison}) and the one from SIMBA's fiducial simulation (green regions in the right panel of Fig. \ref{fig:wind_mass_loading_comparison}) is due to the fact that SIMBA uses \cite{Alcazar2017}'s mass loading trend, which has double the amplitude of the mass loading in \cite{Muratov2015} due to how the two mass loadings are evaluated.

The fiducial ASTRID mass loading values are instead compatible with the IllustrisTNG best-fit simulation values, while the mass loading trend associated to the best-fit analysis is positioned at much lower rotational velocities, and much higher values of the mass loading. This inverted behavior with respect to SIMBA and IllustrisTNG seems to point to some kind of issue with the best-fit simulation detected by our methods, perhaps concerning the lack of a wind velocity floor in equation \eqref{eq:ASTRID_wind_velocity}, which produces very low values at the denominator in equation \eqref{eq:ASTRID_wind_mass_loading_trend}. In fact, the median of the wind velocity distribution for the ASTRID best-fit simulation is $40\,\textrm{km/s}$, while for IllustrisTNG we obtain $350\,\textrm{km/s}$.

We also performed a linear regression (in log-space) of the wind mass loading factor at injection trends, both for the fiducial and the best-fit simulation. Outflows powered by stellar feedback are thought to be driven a) either by momentum, injected into the ISM by massive stellar winds and SNe through radiation pressure, with a power-law scaling $\eta_{w} \propto V_{\textrm{max}}^{-1}$, or b) by energy, injected into the ISM by massive stars and core-collapse SNe, in which case the scaling is $\eta_{w} \propto V_{\textrm{max}}^{-2}$ (see \citealt{Dekel1986, Murray2005, Hopkins2012}).

We obtain for the fiducial simulation:

\begin{equation}
\log_{10}(\eta_{\textrm{w}}) = (-2.36\pm 0.08)\log_{10}(V_{\textrm{max}}) + (5.5\pm 0.2),
\end{equation}
\\
while for the best-fit simulation we obtain:

\begin{equation}
\log_{10}(\eta_{\textrm{w}}) = (-1.99\pm 0.07)\log_{10}(V_{\textrm{max}}) + (4.2\pm 0.1),
\end{equation}
\\
which has a slightly shallower slope than the fiducial trend, closer to the theoretical $V_{\textrm{max}}^{-2}$ trend. These trends are shown as black dashed (fiducial) and dotted (best-fit) curves in Fig. \ref{fig:wind_mass_loading_comparison}. In literature \citep{Muratov2015}, a double power-law trend is used to describe analytically the wind mass loading trend as a function of $V_{\textrm{max}}$. For simplicity, we used a simple power-law, since we do not have many low-velocity galaxies due to our selection criteria.

We performed a linear regression also of the SIMBA and ASTRID wind mass loading factor trends. For SIMBA, we obtain for the fiducial simulation:

\begin{equation}
\log_{10}(\eta_{\textrm{w}}) = (-2.90\pm 0.11)\log_{10}(V_{\textrm{max}}) + (7.6 \pm 0.3),
\end{equation}
\\
while for the best-fit simulation we get:

\begin{equation}
\log_{10}(\eta_{\textrm{w}}) = (-1.27 \pm 0.15)\log_{10}(V_{\textrm{max}}) + (3.1\pm 0.3).
\end{equation}

For ASTRID instead, we obtain for the fiducial simulation:

\begin{equation}
\log_{10}(\eta_{\textrm{w}}) = (-2.11\pm 0.02)\log_{10}(V_{\textrm{max}})+(4.75\pm0.04),
\end{equation}
\\
while for the best-fit simulation we get:

\begin{equation}
\log_{10}(\eta_{\textrm{w}}) = (-2.43\pm0.01)\log_{10}(V_{\textrm{max}}) + (6.25\pm 0.03)
\end{equation}

We caution on the slope values of the SIMBA and ASTRID best fit-simulations, since they are obtained only from galaxies within a tight range of velocities at $V_{\rm max} \lesssim 150 \, \rm km/s$. This is especially true for ASTRID, given the peculiar behavior described above.

\section{Discussion}\label{sec:Discussion}

In this paper we have built a new method for constraining astrophysical and cosmological parameters, by comparing scaling relations predicted by simulations with those constructed with the data. Depending on very specific details of the simulations analyzed, as for example the cosmological parameters and the recipes adopted for the SN feedback, and the selection criteria adopted, we generated samples of simulated galaxies more or less abundant and characterized by a wide range of physical properties. We have optimized the use of this wealth of information to constrain first the astrophysical processes, for instance, wind energy, wind velocity and wind mass loading and the AGN-related parameters, and also cosmological parameters, such as $\Omega_\textrm{m}$, $\sigma_8$ and $S_{8}$.

\subsection{Discrepancy between fiducial simulations and observations}
This analysis has firstly allowed us to highlight a strong discrepancy among the three suites implemented in the \textsc{camels} simulations. In the literature, it is already known that the IllustrisTNG simulation systematically underestimates the SPARC trends in regards to the stellar-to-halo mass relations $M_{*}/M_{\textrm{tot}}$-$M_{*}$ and $M_{*}/M_{\textrm{tot}}$-$M_{\textrm{tot}}$ \citep{Romeo2020}, in agreement with our findings. The comparison between IllustrisTNG, SIMBA and ASTRID fiducial simulations shows that the former aligns more closely to the observed SPARC scaling relations than the latter two, especially for those relations that involve internal quantities. More specifically, SIMBA shows a systematically larger DM mass and DM fraction in the central regions, compared with IllustrisTNG, while ASTRID also shows systematically larger stellar half-mass radii for all values of $M_{*}$, and higher total masses with respect to SPARC for high values of $M_{*}$ ($\sim 10^{10.5}\,M_{\odot}$).

Some discrepancies of SIMBA with observations have been noted in both \cite{Davé2019} and \cite{Glowacki2020}. In the first, it is reported that SIMBA fails to reproduce correctly the stellar mass function at $z=0$, the sizes of quenched low-mass galaxies and the production of stellar metallicity, as well as $\textrm{sSFR}$, in low-mass star-forming galaxies. In the second, it is reported that SIMBA produces galaxies which are overly bulge-dominated, due to the implementation of the feedback from star formation. Moreover, in \cite{Marasco2020}, it is noted that there is a strong discrepancy in the stellar-to-dark matter ratio of simulated to observed systems, which extends into the innermost regions of galaxies.

A possible explanation for this discrepancy could be the fact that galactic winds associated with SN feedback in SIMBA do not interact with gas particles from the ISM, due to hydrodynamic decoupling implemented in the simulation (see \citealt{Alcazar2017}). As reported in \cite{Glowacki2020}, this could lead to overly bulge-dominated star-forming galaxies, with a corresponding overdensity of dark matter particles in the central regions. Another explanation for the discrepancy could be a higher wind mass loading contribution in SIMBA simulations, as shown in the right panel of Fig. \ref{fig:wind_mass_loading_comparison}. Strong baryonic mass ejections from the internal regions in SIMBA, along with effects from hydrodynamical decoupling, could skew the DM fraction evaluated within the stellar half-mass radius towards higher values.

The systematic increase in the central DM mass instead seems to be a long-lasting issue of hydrodynamical simulations \citep{Navarro2000, Marasco2020}, which seems to not be explicable without demanding substantial revisions of the simulated model of structure formations. There could also be an increased effect of adiabatic contraction in galaxies for SIMBA simulations: the fact that baryonic infall drags towards the center of the galaxy the DM particles more intensely in SIMBA simulations than in IllustrisTNG simulations could be a possible explanation of this strong increase in central DM mass \citep{Gnedin2004, Napolitano2010}.

Finally, concerning the ASTRID fiducial simulation, at fixed stellar masses, we find systematically high values for all the physical quantities investigated: galaxies are larger than observations and contain more dark matter. It is not clear yet what could be the reason behind the observed discrepancies with the observed SPARC trends. A more detailed analysis of the physical reasons behind this discrepancy is deferred to future CASCO papers.

\subsection{One-parameter variation results' interpretations}\label{sec:1P_results_interpretations}
To evaluate the impact of SN and AGN feedback on scaling relations, we have also investigated, for a fiducial cosmology, the impact of varying the astrophysical parameters. For the `1P' simulations analysis, the main result is that scaling relation normalization in IllustrisTNG seems to be mainly affected by the supernova feedback parameter $A_{\textrm{SN1}}$, with lower values of the wind energy per unit SFR being associated, at fixed stellar mass, to lower values of stellar half-mass radius, internal DM mass and fractions, total mass and higher stellar masses per galaxy, shifting the correlations to match the observations. These results are not surprising: lower SN feedback implies a higher number of stars formed per galaxy, which implies a higher stellar mass, which is one of the physical driver of the changes in the correlations (see Figs. \ref{fig:selection_effects_sims} and \ref{fig:1P_variation_comparison}). However, more energetic winds are expected to push the gas to larger distances if compared to a weaker feedback, altering the gravitational potential. More/less energetic winds, indeed, increase/reduce half-mass radii, and consequently DM mass and DM fraction.

The impact of the SN feedback parameter $A_{\textrm{SN2}}$ on the correlation between the stellar half-mass radius and stellar mass is inverted, with lower values of $A_{\textrm{SN2}}$ associated with higher stellar half-mass radii, at fixed stellar mass, and a larger number of galaxies at the high-mass end. A possible answer for this could be the fact that, in the definition of wind mass loading, wind speed at injection is at the denominator, so that lower values of $A_{\textrm{SN2}}$ have an opposite effect on $\eta_{\textrm{w}}$ with respect to a decrease in $A_{\textrm{SN1}}$. This means that lower values of wind speeds at injection are associated with a higher amount of galactic outflows, implying higher stellar half-mass radii. The impact on the other correlations is mild.

The analysis on the `1P' simulation set also shows that the effects of varying $A_{\textrm{AGN1}}$ and $A_{\textrm{AGN2}}$ on all the scaling relation trends are negligible. This could be because, in the $[10^{9},\, 10^{11}]\,\textrm{M}_{\odot}$ stellar mass range, AGN feedback effects are weaker with respect to SN feedback. In \cite{Irodotou2022} for example, it is noted that AGN feedback mechanisms mainly influence the star distribution, star formation and gas outflows in the central kiloparsec regions, and only slightly affect the total-stellar mass scaling relation of barred, Milky Way-like galaxies, which lie at the higher end of the mass interval considered in this paper. 

In the case of SIMBA and ASTRID, the figures relative to the 1P analysis have been shown and discussed in detail in Appendix \ref{sec:SIMBA_ASTRID_1P_results}. For the SIMBA suite, considering again the reference cosmological parameters, the dependence of the scaling relations on the wind mass loading is negligible, while some variations are seen with respect to the wind velocity. By combining the wind velocity and the wind mass loading, we also see that an increase of $\tilde{A}_{\textrm{SN1}}$ or  $\tilde{A}_{\textrm{SN2}}$ correspond to larger wind energy.  The wind velocity seems to have a larger impact than in IllustrisTNG, and inverted, with slower winds producing smaller DM and total mass. The large central DM mass, produced by SIMBA in this reference cosmology, could be the possible cause of such independence or small dependence of scaling relations by the SN feedback parameters.

There could also be a saturation effect with respect to the scaling relation trends in the SIMBA simulations at high values of $\Omega_{\textrm{m}}$, preventing the wind mass loading to show its effect on the trends. Indeed, we have verified that lowering the value of $\Omega_{\textrm{m}}$ to $\sim 0.10$ creates a greater variation in scaling relation trends than that which can be seen in Fig. \ref{fig:1P_variation_comparison_SIMBA_ASTRID}. This can also be slightly seen in Fig. \ref{fig:1P_cosmo_variation_comparison_SIMBA_ASTRID}, where we need values of $\Omega_{\textrm{m}}$ close to 0.10 to have a reconciliation with the observations. This potential saturation effect could then explain the low response to the variations of $\tilde{A}_{\textrm{SN1}}$ that we see in Fig. \ref{fig:1P_variation_comparison_SIMBA_ASTRID} for the SIMBA simulations.

For the ASTRID suite, one has to be very careful in tracing the actual effects on the scaling relations of each parameter. For $\hat{A}_{\textrm{SN1}}$, the effects on the scaling relations are similar to the case of IllustrisTNG, but with the difference that lower values of $\hat{A}_{\textrm{SN1}}$ correspond to a lower number of LTG galaxies having high stellar mass. This could be a selection effect: a lower value of $\hat{A}_{\textrm{SN1}}$ could indirectly affect the $\textrm{sSFR}$ in such a way as to convert most of the high stellar mass galaxies into passive galaxies.
For $\hat{A}_{\textrm{SN2}}$, the main effect of decreasing this parameter seems to be an increase in the formation of LTG galaxies with respect to simulations with higher values of the parameter, but also a reduction of the galaxies' stellar masses to around $M_{*}\sim2\times 10^{9}\,M_{\odot}$. A speculative explaination for this effect could come by reading the right panel in figure 4 of \cite{Ni2023}. Here, we see that the star formation rate density (SFRD) in ASTRID is systematically higher at higher redshifts than the other two simulations. This implies that galaxies in ASTRID started to form stars much faster than in IllustrisTNG and SIMBA. For low values of $\hat{A}_{\textrm{SN2}}$ then, gas remains trapped more easily in galaxies due to lower wind outflow velocities, and due to the very high SFRD in their past histories, these galaxies started to convert gas in stars and exhaust cold gas faster than in IllustrisTNG and SIMBA. In the end, at $z=0$ in ASTRID only low stellar mass galaxies will remain with enough gas content left to form stars, which will have a higher $M_{\textrm{g}}/M_{*}$ ratio than the galaxies at high mass. This has been verified by plotting $M_{\textrm{g}}/M_{*}$ for all ASTRID galaxies.

As far as the AGN feedback in ASTRID is concerned, the thermal mode seems to affect the scaling relations more than the kinetic mode. In particular, as detailed in \cite{Ni2023}, a higher value of $\hat{A}_{\textrm{AGN2}}$ is associated to an heightened star-formation, due to a positive feedback induced by the fact that larger values of $\hat{A}_{\textrm{AGN2}}$ suppress the formation of massive black holes, which brings less baryonic suppression on the total matter power spectrum. This is seen in Fig. \ref{fig:1P_variation_comparison_SIMBA_ASTRID}, in that the simulation with $\hat{A}_{\textrm{AGN2}} = 0.25$ has a very low number of LTG galaxies present, which do not present a very high extension in stellar mass, while the simulation with $\hat{A}_{\textrm{AGN2}} = 4.00$ shows LTG galaxies even at $M_{*}\geq 10^{11}\,M_{\odot}$.

We would like to comment that, as reported in \cite{Ni2023}, due to the intricacy of how feedback processes effects are conflated numerically, one should try to view the astrophysical parameters not as the numerical amount of feedback that a simulation manifests with respect to the fiducial simulations, but as the modulation of various processes, that lead to variations in many different physical quantities. For example, in ASTRID the matter power spectrum is sensitive to both $\hat{A}_{\textrm{SN2}}$ and $\hat{A}_{\textrm{AGN2}}$, while the global galaxy properties are mainly driven, indirectly, by the $\hat{A}_{\textrm{SN2}}$ parameter.

\subsection{Discussion on bootstrap procedure results and comparison with literature}
\begin{figure*}
\centering
 \includegraphics[width=\linewidth]{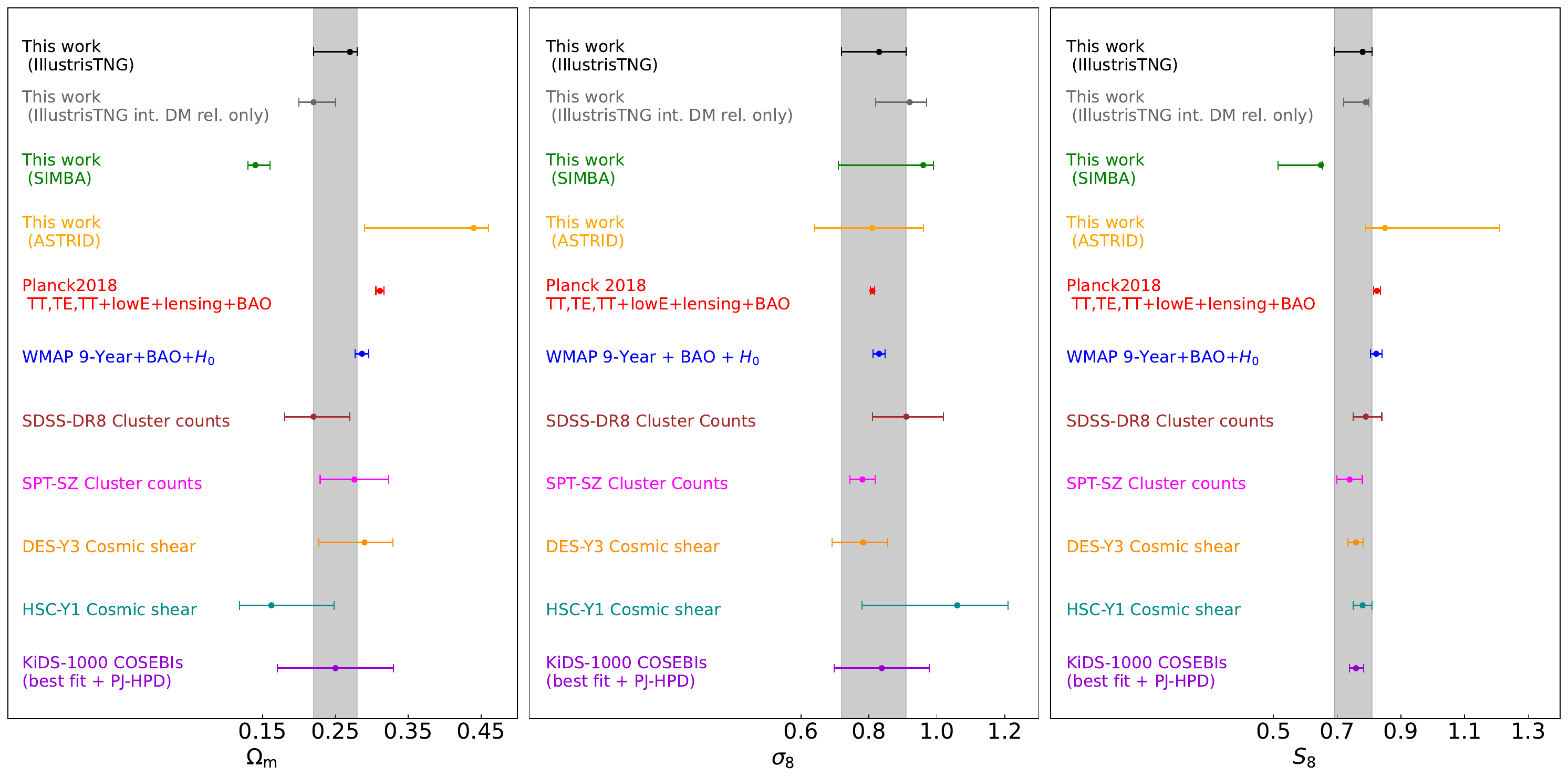}
\caption{Comparison of the constraints on $\Omega_{\textrm{m}}$ (left panel), $\sigma_{8}$ (central panel) and $S_{8}:=\sigma_{8}(\Omega_{\textrm{m}}/0.3)^{0.5}$ obtained from fitting the SPARC star-forming galaxy catalog scaling relation trends with the \textsc{camels}' IllustrisTNG (black point, grey point for DM scaling relations only), SIMBA (green point) and ASTRID (orange point) simulation suites, with results presented, from top to bottom, in \protect\cite{Planck2018} (red point), \protect\cite{Hinshaw2013} (blue point), \protect\cite{Costanzi2019} (brown point), \protect\cite{Bocquet2019} (magenta point), \protect\cite{Amon2022} and \protect\cite{Secco2022} (dark orange point), \protect\cite{Hikage2019} (cyan point) and \protect\cite{Asgari2021} (violet point).  The confidence interval between the 16th and 84th percentile in our measurements is also shown as transparent grey bands.}
\label{fig:cosmo_params_comparison}
\end{figure*}

We have also developed a method to quantify the agreement between simulations and data, by performing a $\chi^2$ minimization. We have shown that the constraining power of the analyzed scaling relations is stronger on $\Omega_{\rm m}$, while the dependence on $\sigma_{\rm 8}$ is milder. However, it is vital for our approach to check the consistency with independent and more robust cosmological parameter probes. In fact, our best-fitted results for the IllustrisTNG suite are in good agreement with almost all the cosmological results presented in literature, as shown in Fig. \ref{fig:cosmo_params_comparison}. We constrain the cosmological parameters with an average precision of 10 per cent, and the quite good agreement with results based on cosmological probes gives credibility to our results and to the constraints on the astrophysical parameters. While the errors on $\Omega_{\textrm{m}}$ are very small (11 percent), in perfect agreement with the estimates obtained using IllustrisTNG, the uncertainty for $\sigma_8$ is of $\sim 15$ per cent, with a predominant tail towards lower values. SIMBA shows an agreement within $1\sigma$ with all literature measurements only for $\sigma_{8}$, while there is agreement only with the $\Omega_{\textrm{m}}$ result presented in \cite{Hikage2019}. ASTRID, finally, is in agreement within $1\sigma$ with the \cite{Planck2018} results for all three cosmological parameters, albeit with large error bars that are strongly skewed towards low values of $\Omega_{\textrm{m}}$ and very high values of $S_{8}$.

Regarding the astrophysical parameters, we find that IllustrisTNG shows a constraint of $A_{\textrm{SN1}}$ which is in tension with the fiducial unit value by more than $2\sigma$, which directly indicates that the mass loading of the IllustrisTNG simulations must be lowered to allow compatibility with the observations. This is consistent with the findings of \cite{Yongseok2023}, who find a bimodal posterior distribution for $A_{\textrm{SN1}}$ for which the highest peak is below unity, near $0.50$. The value of $A_{\textrm{SN2}}$ is instead compatible with the fiducial value within $1\sigma$. Due to the fact that AGN feedback processes in IllustrisTNG have a negligible effect on the scaling relations, we find that it is not possible to constrain the parameters $A_{\textrm{AGN1}}$ and $A_{\textrm{AGN2}}$. The results obtained are in agreement with those derived via machine learning approach applied on single galaxies in \cite{Villaescusa2022} and \cite{Echeverri2023}.

In the case of the SIMBA suite instead, we have a match with observations only for unreasonably low $\Omega_\textrm{m}$ ($=0.14$) and high $\sigma_8$ ($= 0.96$) values of the cosmological parameters. This tendency towards having extreme values of the cosmological parameters could be an effect of the potential saturation at high values of $\Omega_{\textrm{m}}$ in SIMBA simulations that we described before, in that we need low values of $\Omega_{\textrm{m}}$ first to be able to break the saturation, and then a variation in astrophysical parameters at fixed (low) values of $\Omega_{\textrm{m}}$ to better fit the observations.

Regarding the astrophysical parameters, we find in the case of SIMBA that the AGN feedback parameters $\tilde{A}_{\textrm{AGN1}}$ and $\tilde{A}_{\textrm{AGN2}}$ are better constrained, while we cannot constrain the $\tilde{A}_{\textrm{SN2}}$ parameter. As discussed in Section \ref{sec:CAMELS_SIMBA_best_fit_analysis}, this result can be explained considering that, for cosmologies with lower values of $\Omega_{\textrm{m}}$, the scaling relations are more strongly affected by AGN feedback.

Finally, in the case of the ASTRID suite, we have a match with observations only with unrealistic mass distributions, where all galaxies are concentrated around $M_{*}\sim2\times 10^{9}\,M_{\odot}$. This is because the $\chi^{2}$ minimization finds the simulations with the best trends, and in ASTRID all simulations which have LTGs at high stellar mass do not reproduce the observed relations as well as the ones with clustered galaxies around the median. These simulations all have low values of $\hat{A}_{\textrm{SN2}}$, $\hat{A}_{\textrm{AGN2}}$ and high values of $\Omega_{\textrm{m}}$, giving an indication that these parameters are primarily responsible for this behavior in the simulations. 

Overall, the best constraints seem to come from the IllustrisTNG suite, which does not show the problems that SIMBA and ASTRID manifested during this analysis.

\subsection{Wind mass loading discussion}
Regarding the mass loading analysis, to explain the discrepancies between the simulations and the literature results shown in the left panel of Fig. \ref{fig:wind_mass_loading_comparison}, one must first distinguish the different approaches with which mass loading factors are considered in literature.

As reported in \cite{Belfiore2019}, the first approach considers the so-called `mass loading factor at injection', which means that the state of the outflowing gas is directly related to an ongoing star formation event. This is the approach that is also used in hydrodynamical simulations which use a sub-grid for launching winds, such as \textsc{camels} (see \citealt{Pillepich2018}). The second one considers a `time-averaged cumulative mass loading factor', which is the ratio between the star formation rate and the amount of gas leaving the galaxy's halo over a defined time-scale (see \citealt{Muratov2015}).
Usually, the cumulative mass loading factor is up to an order of magnitude lower than the instantaneous loading factor, which could explain why the results from \cite{Peeples2011}, \cite{Lilly2013} and \cite{Zahid2014} are systematically lower than both our results and the hydrodynamical simulations' results from \cite{Davé2011} and \cite{Muratov2015}. Given that the empirically determined values depend on the metallicity calibrations and oxygen nucleosynthetic yields, changing these two parameters in the observations could give higher loading factors than the ones shown in Fig. \ref{fig:wind_mass_loading_comparison}.

As far as the mass loading trends from the simulations are concerned, in both IllustrisTNG and SIMBA suites there is a tendency for the best-fit simulations to decrease their mass loading values with respect to the corresponding fiducial simulations, which brings them closer to the hydrodynamical simulation results from \cite{Davé2011} and \cite{Muratov2015}. Only in the case of the ASTRID simulation we observe the reverse, in that the best-fit simulation has unusually high mass loading values and low maximum rotation velocities compared to the fiducial simulation, which is instead closer to the literature results. Along with the unrealistic mass distribution discussed previously, this result for the mass loading in ASTRID reinforces the idea that the simulation that better reproduces the SPARC trends from ASTRID is physically unrealistic.

\section{Conclusions}\label{sec:Conclusions}

In this work, we have introduced the project {\it CASCO: Cosmological and AStrophysical parameters from Cosmological simulations and Observations}, which aims at comparing simulations and observations for constraining cosmological parameters and astrophysical processes.

In this first paper of the series, we compare various scaling relations for star-forming galaxies, taken from the IllustrisTNG, SIMBA and ASTRID subgrid-based suites of the \textsc{camels} simulations \citep{Villaescusa2021}, with observed data from the star-forming galaxy catalog SPARC \citep{Lelli2016}. The simulated sample consists, for each simulation, of all those galaxies having $R_{*,1/2} > \epsilon_{\textrm{min}}$, $N_{*,1/2} > 50$ and $f_{\textrm{DM}} (<R_{*,1/2}) > 0$, while the observed SPARC sample is made up by 152 star-forming galaxies, binned with respect to the stellar mass. The scaling relations considered are the size-mass relation ($R_{*,1/2}$-$M_{*}$), the internal DM fraction against stellar mass ($f_{\textrm{DM}}(<R_{*,1/2})$-$M_{*}$), the internal DM mass against stellar mass ($M_{\textrm{DM},1/2}$-$M_{*}$) and the total-stellar mass relation ($M_{\textrm{tot}}$-$M_{*}$). 

\begin{itemize}
\item We started by comparing the fiducial simulations ($\Omega_{\textrm{m}} = 0.30$, $\sigma_{8} = 0.80$, $A_{\textrm{SN1}} = A_{\textrm{AGN1}} = A_{\textrm{SN2}} = A_{\textrm{AGN2}} = 1.00$) of the three simulation suites. IllustrisTNG shows a better agreement with the observed scaling relation trends, especially in regards to trends involving internal quantities, e.g. $M_{\textrm{DM},1/2}$, with a cumulative (i.e. sum of all the contributions from the single scaling relations) normalized chi-squared of $\tilde{\chi}^{2} = 2.33$ for IllustrisTNG, against $\tilde{\chi}^{2} = 6.20$ for the SIMBA fiducial simulation and $\tilde{\chi}^{2} = 8.49$ for the ASTRID fiducial simulation.

\item We then proceeded by varying the two cosmological parameters, $\Omega_{\textrm{m}}$ and $\sigma_{8}$, and the four astrophysical parameters, $A_{\textrm{SN1}}$, $A_{\textrm{SN2}}$, $A_{\textrm{AGN1}}$ and $A_{\textrm{AGN2}}$, which regulate the SN feedback and the AGN feedback processes, respectively, one by one. We varied each of the six parameters between the minimum and the maximum of the allowed range, and compared the resulting simulated trends to the observed trends from SPARC. Results show that simulations with a lower value of the astrophysical parameter $A_{\textrm{SN1}}$ better reproduce the observed trends in all three simulation suites, while strong variations of both AGN feedback parameters in the IllustrisTNG simulation suite show negligible effects on the scaling relations considered. This is not surprising, since the role of the AGN feedback is expected to be more relevant in more massive galaxies. On the other hand, by fixing the cosmological parameters to the reference values, SIMBA simulations predict scaling relations which do not depend on wind mass loading and AGN parameters, and show a dependence only from the wind velocity. These small dependencies, and the systematically high central DM mass produced in the reference cosmology, necessarily require a change in the cosmological parameters in order to accommodate the observations. Finally, ASTRID simulations show a weak dependency on the wind mass loading, in a way similar to the case of the IllustrisTNG suite, and show peculiar clustering effects at low values of $\hat{A}_{\textrm{SN2}}$. While there is still no dependency on the AGN parameter $\hat{A}_{\textrm{AGN1}}$, which regulates the kinetic AGN feedback mode, there is some dependency on $\hat{A}_{\textrm{AGN2}}$, the parameter which regulates the thermal AGN feedback mode, in that higher values of this parameter enhance star-formation in galaxies due to a positive feedback regarding the suppression of the formation of massive black holes \citep{Ni2023}.

\item We next considered all 1065 simulations of the `LH', `1P' and `EX' sets in the IllustrisTNG suite, performed a bootstrap resampling 100 times on both the simulation points and the SPARC dataset, and searched for the best-fit simulation associated to each resampling, in order to obtain constraints on the cosmological and astrophysical parameters by considering the parameter distributions associated to the best-fit simulations. We obtain $\Omega_{\textrm{m}} = 0.27_{-0.05}^{+0.01}$, $\sigma_{8} = 0.83_{-0.11}^{+0.08}$, $S_{8} = 0.78_{-0.09}^{+0.03}$, $A_{\textrm{SN1}} = 0.48_{-0.16}^{+0.25}$, $A_{\textrm{SN2}} = 1.21_{-0.34}^{+0.03}$, $A_{\textrm{AGN1}} = 2.53_{-1.82}^{+0.89}$ and $A_{\textrm{AGN2}} = 1.31_{-0.67}^{+0.49}$ with IllustrisTNG, $\Omega_{\textrm{m}} = 0.14_{-0.01}^{+0.02}$, $\sigma_{8} = 0.96_{-0.25}^{+0.03}$, $S_{8} = 0.649_{-0.135}^{+0.004}$, $\tilde{A}_{\textrm{SN1}} = 0.45_{-0.10}^{+0.06}$, $\tilde{A}_{\textrm{SN2}} = 0.81_{-0.31}^{+1.03}$, $\tilde{A}_{\textrm{AGN1}} = 0.56_{-0.02}^{+0.12}$ and $\tilde{A}_{\textrm{AGN2}} = 1.13_{-0.35}^{+0.03}$ with SIMBA and, finally, $\Omega_{\textrm{m}} = 0.44_{-0.15}^{+0.02}$, $\sigma_{8} = 0.81_{-0.17}^{+0.15}$, $S_{8} = 0.85_{-0.06}^{+0.36}$, $\hat{A}_{\textrm{SN1}} = 0.41_{-0.17}^{+0.34}$, $\hat{A}_{\textrm{SN2}} = 0.61_{-0.04}^{+0.12}$, $\hat{A}_{\textrm{AGN1}} = 2.49_{-0.90}^{+0.56}$ and $\hat{A}_{\textrm{AGN2}} = 0.62_{-0.09}^{+1.05}$ with ASTRID. We thus manage to constrain $\Omega_{\textrm{m}}$ and $A_{\textrm{SN1}}$ with good precision, while parameters $\sigma_{8}$, $S_{8}$ and $A_{\textrm{SN2}}$ are constrained with lower precision with IllustrisTNG. Parameters $A_{\textrm{AGN1}}$ and $A_{\textrm{AGN2}}$ are instead not constrained. This is in agreement with other results using \textsc{camels} \citep{Villaescusa2022, Echeverri2023}. Cosmological constraints obtained from the IllustrisTNG suite are consistent within $1\sigma$ with many measurements reported in literature, in particular with results from \cite{Hinshaw2013}. In contrast, constraints from SIMBA are only in agreement with the literature when considering the $\sigma_{8}$ estimate, while those from ASTRID are primarily in agreement with the results from \cite{Planck2018}, as illustrated in Fig. \ref{fig:cosmo_params_comparison}.
Astrophysical constraints in all three simulations seem to agree that a lower value of $A_{\textrm{SN1}}$ (which regulates wind outflow energy per unit SFR in IllustrisTNG, and more broadly the global wind mass loading trend in SIMBA and ASTRID) than the fiducial unit value are needed in order to be consistent with the observations from SPARC. This is also in agreement with results from \cite{Yongseok2023}, whose posterior result for $A_{\textrm{SN1}}$ show a bimodal distribution, linked to degeneracies that exist between the cosmological and astrophysical parameters of \textsc{camels}, which peaks more strongly around values close to $0.50$ instead of the reference unitary value.

\item Finally, by analyzing the wind mass loading trends, we find that in IllustrisTNG both the fiducial and the best-fit simulation provides a good agreement with other hydrodynamical simulation results, e.g. \cite{Davé2011} and \cite{Muratov2015}, with the best-fit simulation having a slope compatible with the one theoretically expected for energy-driven winds. The observable discrepancies with empirical estimates of the wind mass loading factor from \cite{Peeples2011}, \cite{Lilly2013} and \cite{Zahid2014} arise because we are considering mass loading factors at injection instead of time-averaged cumulative loading factors, which are usually an order of magnitude lower than the former. There could also be a systematic effect due to metallicity calibration and nucleosynthetic yields choice that influences the empirical trends. SIMBA mass loading trends manage to reconcile with observations only by considering the best-fit simulation, i.e. lowering $\Omega_{\textrm{m}}$ to values near $0.10$, while the fiducial ASTRID simulation is located near the trends of \cite{Muratov2015}, between the fiducial and the best-fit IllustrisTNG trends, and also shows a slope compatible with energy-driven winds. The best-fit ASTRID simulation shows, instead, a strong discrepancy with the literature, being located at higher values than the SIMBA fiducial simulation and clustered around very low rotational velocity values.
\end{itemize}

In this paper we have started to test the predictive power of galaxy scaling relations, by comparing hydrodynamical simulations and observations. We have limited our analysis to study the size, central and total dark matter content in local star-forming galaxies. Probing the evolution of such correlations with cosmic time \citep{Sharma2022} will be the next step, to trace back in cosmic time and in more detail the physical processes underlying the scaling relations. We expect, in fact, that physical properties of galaxies show a stronger dependence on both cosmology and astrophysical parameters at larger redshifts (see, e.g., the star formation rate density in Fig. 9 of \citealt{Villaescusa2021}). This approach could also be used for calibrating simulations with respect to observations, by finding the parameter models that better fit a certain observational dataset. In future works, we will also perform a more sophisticated statistical analysis, as far as the comparison between the observed and the simulated scaling relations is concerned.
Next generation radio surveys, such as the Widefield ASKAP L-band Legacy All-sky Blind surveY (WALLABY, \citealt{Koribalski2020}), will also allow more precise measurements of rotation curves by studying the HI properties of galaxies, which will provide lower scatters associated to the observed scaling relation trends. To have a more complete view in terms of mass and galaxy types, we plan on applying the same procedure to passive galaxies in the future, e.g. comparing with local galaxies (SPIDER, \citealt{LaBarbera2010, Tortora2012}), as well as studying the redshift evolution of the respective scaling relations \citep{Tortora2014,Tortora2018}. Strong lenses also gives an important constraining tool for the galaxy processes \citep{Koopmans+06_SLACSIII, Gavazzi+07_SLACSIV,Auger+10_SLACSX,Tortora+10lensing}, and the future Euclid mission or Rubin/LSST will provide hundred thousands lenses which, complemented with spectroscopic information, will provide unprecedented constraints on the masses and mass profiles of massive galaxies.

\section*{Acknowledgements}

All the calculations underlying this work have been performed via the use of Wolfram Mathematica ver. 13.1. V.B., C.T. and F.G. acknowledge the INAF grant 2022 LEMON. We thank Francisco Villaescusa-Navarro, Daniel Anglés-Alcázar and Shy Genel for their advice. We thank the anonymous referee for his/her comments, which helped us to better present our results.

\section*{Data Availability}

The data underlying this article will be shared on reasonable request to the corresponding author.



\bibliographystyle{mnras}
\bibliography{bibliography}




\appendix
\section{Systematics}
Below we will discuss various forms of systematic errors that could affect the final results of this paper, and their impact on the latter.

\subsection{Constant M/L ratio and projected-3D radii conversion contributes}\label{sec:const_ML_ratio_and_2D_3D_conversion}

In Section \ref{sec:SPARC_data}, we fixed the stellar mass-to-light ratio at $3.6\;\mu\textrm{m}$ to the value of $\Upsilon_{*} = 0.6\;\Upsilon_{\odot}$. To check for possible systematic effects on the astrophysical and cosmological parameters due to this choice, we repeated the analysis by changing the M/L ratio between $\Upsilon_{*} = 0.5\;\Upsilon_{\odot}$ and $\Upsilon_{*} = 0.7\;\Upsilon_{\odot}$. The impact on stellar mass values is around $0.06$-$0.08\;\textrm{dex}$, while the impact on the final results, obtained by repeating the bootstrap procedure with the new stellar mass values, is negligible (the variation of the parameters is within the uncertainties reported for the bootstrap analysis).

In the same section, we reported that the effective radii of the SPARC galaxies, which are projected quantities, had to be converted into stellar half-mass radii, which are 3D quantities, so as to make a fair comparison. To perform the comparison we used the following relation between $R_{*,1/2}$ and $R_{\textrm{e}}$, valid for Sérsic profiles (\citealt{Wolf2010}, Appendix B):

\begin{equation}
R_{*,1/2}/R_{\textrm{e}} = 1.3560 -0.0293\,n^{-1}+0.0023\,n^{-2},
\end{equation}
where $n$ is the Sérsic index. This equation has, in general, 0.25 per cent accuracy after testing against the numerical integration of Sérsic profiles with $n\in[0.10,2.0]$. In our analysis, we truncated this relation to the first term, which \cite{Wolf2010} reports to be an accurate approximation to better than 2 per cent for most surface brightness profiles used to describe galaxies. We thus consider de-projection systematics due to conversion of physical quantities to be negligible.

\subsection{Binning procedure contribution}\label{sec:binning_procedure_contribution}
As discussed in Section \ref{sec:sim_base_comparison}, we binned the SPARC galaxies in fixed bins of stellar mass. To evaluate the effect that changing the bin edges has on the analysis, we have varied the position of the edges by shifting them randomly in the interval $[-0.2,0.2]\;\textrm{dex}$, and repeated this procedure $N=100$ times. The final results are unchanged with respect to using the reference SPARC scaling relation trends. More specifically, the uncertainty of the median trend for the log-spaced quantities is no higher than $0.09\;\textrm{dex}$, while for the internal DM fraction the relative percent uncertainty is no higher than 5 per cent. The reference median is within the uncertainty intervals for all scaling relation trends. Because of these results, we consider the impact of having fixed the bins in the SPARC binning procedure to be negligible.

\subsection{Cosmic variance contribution}\label{sec:cosmic_variance}
To understand the weight that cosmic variance has on the quantities that enter the scaling relations we want to analyze, we considered the 27 CV simulations from all \textsc{camels} suites. For each CV simulation, we evaluated the median of the following quantities: stellar mass within the stellar half-mass radius, $M_{*,1/2}$, DM mass within the stellar half-mass radius, $M_{\textrm{DM,1/2}}$, total stellar mass, $M_{*}$, total mass, $M_{\textrm{tot}}$, defined in \textsc{camels} as the sum of the mass of all the particle types bound to a subhalo, stellar half-mass radius, $R_{*,1/2}$, DM fraction within the stellar half-mass radius ($f_{\textrm{DM}}(<R_{*,1/2})$), specific star formation rate ($\textrm{sSFR}$) and number of star particles within the stellar half-mass radius ($N_{*,1/2}$).

The standard deviation for each of the parameters in both \textsc{camels} suites are reported in Table \ref{tab:CV_analysis_confidence_intervals}. All the parameters show values of $\sigma$ around $10^{-2}$. We thus consider the contribution due to cosmic variance on the parameters' uncertainty to be negligible.

\begin{table}
\centering

\caption{Standard deviation for various galaxy physical properties due to cosmic variance.}

\begin{tabular}{cccc}
\hline
\hline
quantity & $\sigma$ (IllustrisTNG) & $\sigma$ (SIMBA) & $\sigma$ (ASTRID)\\
\hline
$\log_{10}(M_{*,1/2})$ & $2.7\times 10^{-2}$ & $4.5\times 10^{-2}$ & $3.8\times 10^{-2}$\\
$\log_{10}(M_{\textrm{DM,1/2}})$ & $1.7\times 10^{-2}$ & $1.9\times 10^{-2}$ & $1.7 \times 10^{-2}$\\
$\log_{10}(M_{*})$ & $2.8\times 10^{-2}$ & $4.4\times 10^{-2}$& $3.8\times 10^{-2}$\\
$\log_{10}(M_{\textrm{tot}})$ & $2.2\times 10^{-2}$ & $4.9\times 10^{-2}$& $3.5\times 10^{-2}$\\
$\log_{10}(R_{*,1/2})$ & $1.3\times 10^{-2}$ & $9.4\times 10^{-3}$& $1.2\times 10^{-2}$\\
$f_{\textrm{DM}}(<R_{*,1/2})$ & $5.9\times 10^{-3}$ & $2.6\times 10^{-3}$& $6.6\times 10^{-3}$\\
$\log_{10}(\textrm{sSFR})$ & $3.9\times 10^{-2}$ & $3.3\times 10^{-2}$& $3.1\times 10^{-2}$\\
$N_{*,1/2}$ & $1.0\times 10^{1}$ & $1.7\times 10^{1}$& $1.7\times 10^{1}$\\
\hline
\end{tabular}
    \label{tab:CV_analysis_confidence_intervals}
\end{table}

\subsection{Fixed sSFR threshold contribution}\label{sec:fixed_sSFR_contribution}
We have also checked the impact of the fixed $\textrm{sSFR}$ threshold selection on the simulated data. We varied the $\textrm{sSFR}$ threshold by $\pm 0.5$, and evaluated the median of the distributions of the following quantities (pertaining to the fiducial simulation `1P\_1\_0'): $M_{*}$, $R_{*,1/2}$, $f_{\textrm{DM}}(<R_{*,1/2})$, $M_{\textrm{DM,1/2}}$ and $M_{\textrm{tot}}$. The relative discrepancies between using $\log_{10}(\textrm{sSFR}/\textrm{yr}^{-1})=-11$ and $\log_{10}(\textrm{sSFR}/\textrm{yr}^{-1})=-10$ with respect to $\log_{10}(\textrm{sSFR}/\textrm{yr}^{-1})=-10.5$ are always lower than $10^{-3}\,\textrm{dex}$, so we consider that the simulated scaling relations used are not affected by our choice of $\textrm{sSFR}$ threshold.

\section{Variation of cosmological and astrophysical parameters in SIMBA and ASTRID}\label{sec:SIMBA_ASTRID_1P_results}

\begin{figure*}
\includegraphics[width=\linewidth]{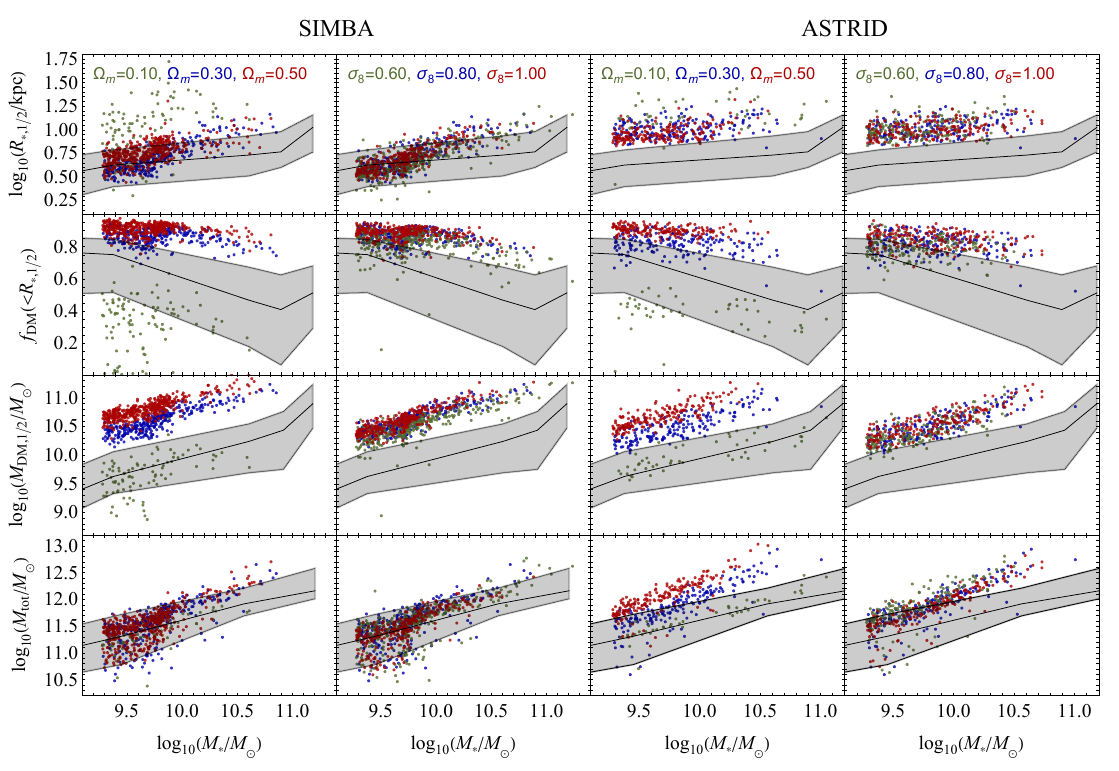}
\caption{Same as Fig. \ref{fig:1P_cosmo_variation_comparison}, but for SIMBA (left panel) and ASTRID (right panel) simulation suites.}
\label{fig:1P_cosmo_variation_comparison_SIMBA_ASTRID}
\end{figure*}

\begin{figure*}
    \centering
    \includegraphics[width=\linewidth]{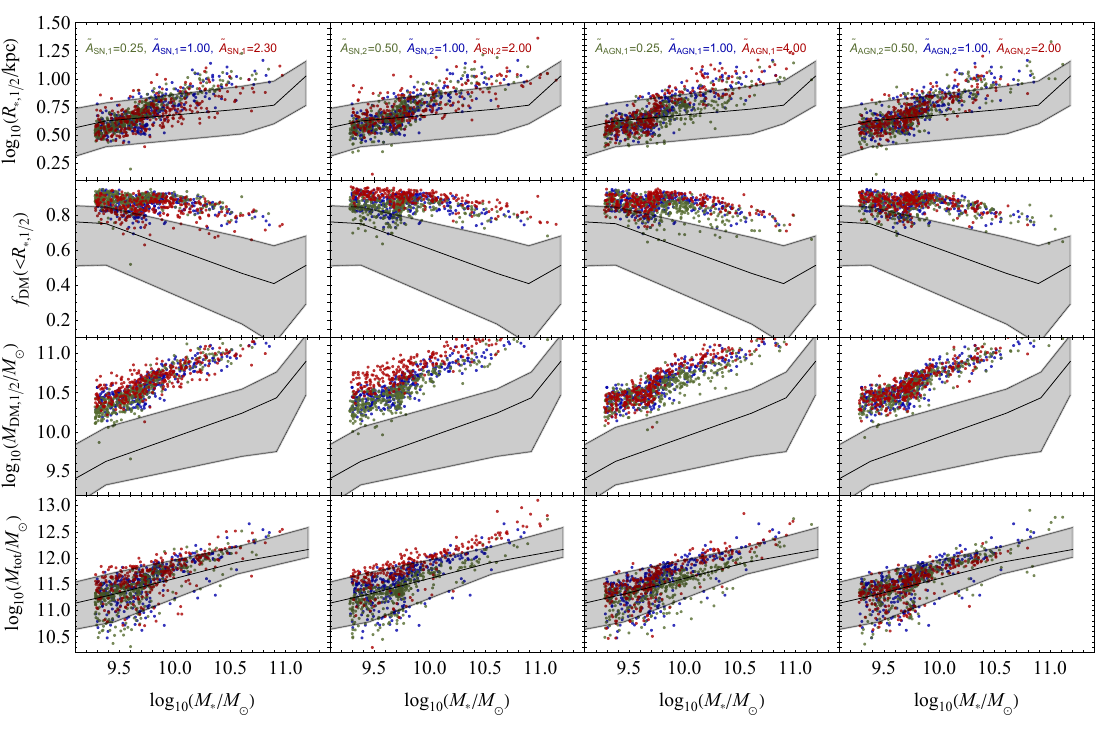}
\includegraphics[width=\linewidth]{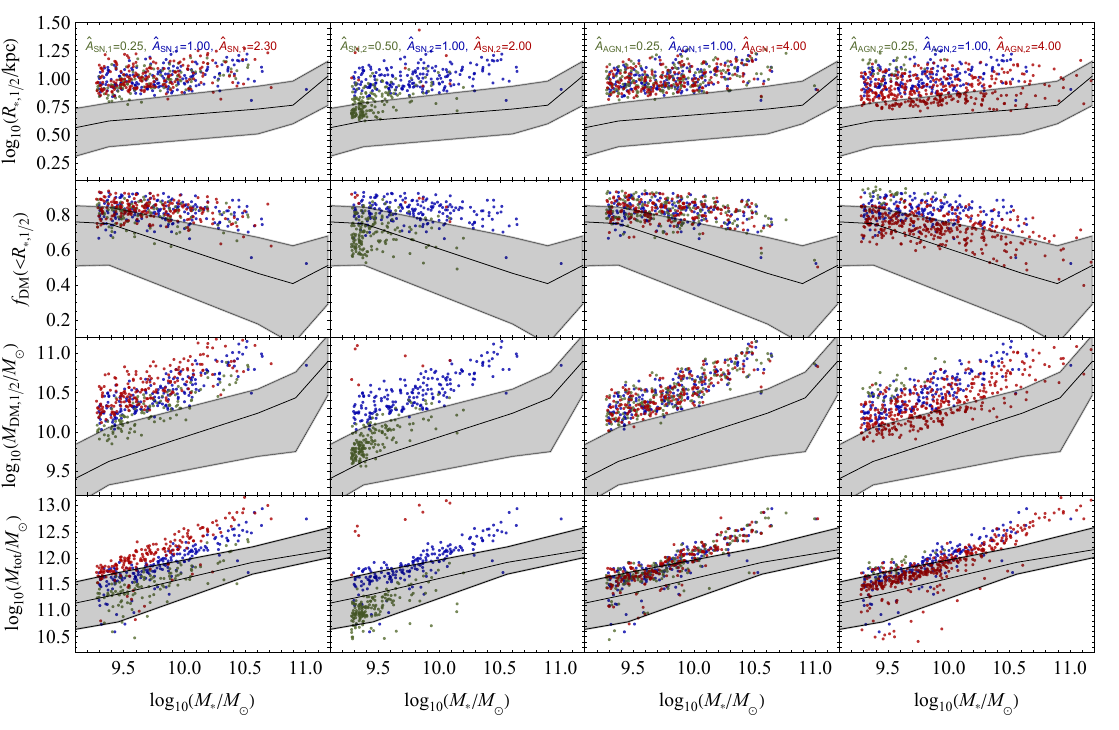}
\caption{Same as Figure \ref{fig:1P_variation_comparison}, but for SIMBA (top panel) and ASTRID (bottom panel) simulation suites.}
    \label{fig:1P_variation_comparison_SIMBA_ASTRID}
\end{figure*}

An identical analysis as the one reported in Section \ref{sec:sim_variable_parameters} has been performed on both SIMBA and ASTRID simulations. Fig. \ref{fig:1P_cosmo_variation_comparison_SIMBA_ASTRID} shows the effects of varying one of the two cosmological parameters on the simulations' trends, while Fig. \ref{fig:1P_variation_comparison_SIMBA_ASTRID} shows the effects of varying one of the four astrophysical parameters on the simulations' trends. In both SIMBA and ASTRID we see similar trends as the ones shown for IllustrisTNG in Fig. \ref{fig:1P_cosmo_variation_comparison}. In both cases, an increase in $\Omega_{\textrm{m}}$ corresponds to an increase in the normalization of the scaling relations. In particular, both in SIMBA and ASTRID there is a slightly better concordance with the observations for values of $\Omega_{\textrm{m}}$ close to $0.10$, especially regarding the $M_{\textrm{DM},1/2}$-$M_{*}$ scaling relation, but in both cases there is also a strong decrease in the number of LTGs present in the simulation. Neither suite shows significant sensitivity to variations of the $\sigma_{8}$ parameter.

As far as the dependency on the astrophysical parameters at fixed cosmological parameters is concerned, the scaling relations in SIMBA are mildly dependent on the SN- and AGN-related parameters. In particular, while the dependence on $\tilde{A}_{\textrm{SN1}}$, which regulates the normalization of the wind mass loading, is negligible, lower velocity winds (lower values of $\tilde{A}_{\textrm{SN2}}$) induce a small decrease in DM fraction, DM mass within the half-mass radius and total mass. The dependence on the AGN parameters seems to also be negligible, with only minor changes induced among the two extreme values used for $\tilde{A}_{\textrm{AGN1}}$.

The dependency of the scaling relations on the astrophysical parameters with fixed cosmological parameters is more complicated in the case of the ASTRID simulations. Similarly to the IllustrisTNG results in Section \ref{sec:sim_variable_parameters}, for an increase in $\hat{A}_{\textrm{SN1}}$ we still see an increase, at fixed stellar mass, in the dark matter mass within the half-mass radius and total mass, but we see almost no effect on the stellar half-mass radius or on the dark matter fraction within the stellar half-mass radius. Moreover, in contrast to the case of IllustrisTNG, in ASTRID there is no increase in stellar half-mass radius with lower values of $\hat{A}_{\textrm{SN2}}$: the trends instead show that simulations with higher $\hat{A}_{\textrm{SN2}}$ possess much more dark matter mass within the stellar half-mass radius, and the latter is also bigger at fixed stellar mass. Moreover, we can see that for low values of $\hat{A}_{\textrm{SN2}}$, galaxies tend to cluster abnormally around $M_{*}\sim 2\times 10^{9}\,M_{\odot}$. There is also a much stronger suppression of star-formation for $\hat{A}_{\textrm{SN2}}>1.00$, as shown by the fact that there are very few LTG galaxies in the simulation with $\hat{A}_{\textrm{SN2}} = 2.00$.

The dependence on the $\hat{A}_{\textrm{AGN1}}$ parameter seems to also be negligible in ASTRID, while there is some variation between simulations with different values of $\hat{A}_{\textrm{AGN2}}$, with higher values corresponding to an enhanced star-formation for high-mass galaxies and smaller half-mass radii. This seems to be consistent with what is reported in \cite{Ni2023}, where in the summary they conclude that a larger $\hat{A}_{\textrm{AGN2}}$ enhances the global star-formation rate via positive feedback mechanisms.

This analysis shows that no combination concerning only astrophysical parameters can reconcile SIMBA and ASTRID central DM masses and DM fractions with observations. Indeed, only forcing the cosmological parameters to extreme values, e.g. reducing $\Omega_\textrm{m}$ to $0.2$ or below, can solve the observed discrepancy of SIMBA with the observations, while for ASTRID the discrepancy can be solved by reducing $\hat{A}_{\textrm{SN2}}$ , but at the cost of having all galaxies clustered at low mass values.

\section{Consistency test for the bootstrap analysis}\label{sec:consistency test}

We performed a consistency test of the bootstrap procedure, to check if the method correctly recovers the cosmological and astrophysical parameters of a given ground-truth, within a certain confidence interval. In the four tests we performed, the ground truth was first the fiducial IllustrisTNG `1P\_1\_0' simulation, then the IllustrisTNG simulation LH-698 (which is also the IllustrisTNG `best-fit' simulation) and, finally, two random simulations taken from the IllustrisTNG LH set. We considered as the mock observational sample a galaxy sample extracted from the ground truth, obtained by randomly extracting 50 per cent of the total number of galaxies of the ground-truth simulation. We then performed the same bootstrapping procedure detailed in Section \ref{sec:CAMELS_best_fit_analysis}. With this procedure, we expect that, after obtaining the list of 100 best-fit simulations, the associated constraints will consistently recover the value of the ground-truth cosmological and astrophysical parameters for each of the ground-truth simulations, within the related uncertainties. The results for each of the six tests are shown in Tables \ref{tab:consistency_test_bestfit}-\ref{tab:consistency_test_LH_25}. For all the four tests, the ground truth is recovered with the cumulative results within $1\sigma$ for all the parameters, with only a few exceptions ($A_{\textrm{SN2}}$ and $A_{\textrm{AGN2}}$ in the third test and the AGN feedback parameters in the fourth test). The parameters that are better constrained are $\Omega_{\textrm{m}}$ and $A_{\textrm{SN1}}$, while on the other hand $\sigma_{8}$. $S_{8}$  and $A_{\textrm{SN2}}$ show slightly higher uncertainties. The parameters $A_{\textrm{AGN1}}$ and $A_{\textrm{AGN2}}$, finally, show consistently high uncertainties, confirming what we already discussed in the main text, i.e. that these parameters are completely unconstrained. We also find that we cannot always recover the ground truth only by using single correlations. We need a larger set of parameters to constrain the cosmological and astrophysical parameters, confirming the goodness of the approach we have followed in this paper. Overall, this test shows that the bootstrap method approach consistently finds lower/higher values for the cosmological and astrophysical parameter, whenever the ground truth is effectively lower/higher than the fiducial results. We also checked what is the effect of observational realism on the consistency test results, by injecting some Gaussian noise according to realistic observable uncertainties into the physical quantities of the ground-truth simulation and running the tests again. Results show that the estimated parameters once again recover the ground truth, but with an increase in the uncertainties. We finally checked what happens if, instead of constraining all the four scaling relations at once, we only constrain the $f_{\textrm{DM}}(<R_{*,1/2})$-$M_{*}$ and the $M_{\textrm{DM,1/2}}$-$M_{*}$  relations (column `cumulative alt.' in Tables \ref{tab:consistency_test_bestfit}-\ref{tab:consistency_test_LH_25}). Results appear to be intermediate between those obtained by constraining only one scaling relation and the cumulative results.

\begin{table*}
\centering
\renewcommand*{\arraystretch}{1.50}
\caption{Constraints on cosmological and astrophyiscal parameters, based on the methods detailed in Section \ref{sec:CAMELS_best_fit_analysis}, obtained by considering a mock observational sample taken from the fiducial IllustrisTNG `1P\_1\_0' simulation, considered as the ground-truth simulation. The constraints are given in terms of 16th, 50th and 84th percentiles. The `cumulative alt.' column values are the cumulative chi-squared results obtained by considering only the internal dark matter scaling relations.}

\begin{tabular}{cccccccc}
\hline
\hline
Parameter & $R_{*,1/2}$-$M_{*}$ & $f_{\textrm{DM}}(<R_{*,1/2})$-$M_{*}$ & $M_{\textrm{DM},1/2}$-$M_{*}$ & $M_{\textrm{tot}}$-$M_{*}$ & cumulative & cumulative alt. & ground-truth\\
\hline 
$\Omega_{\textrm{m}}$ & $0.30_{-0.06}^{+0.05}$ & $0.26_{-0.01}^{+0.03}$ & $0.30_{-0.04}^{+0.03}$ & $0.21_{-0.06}^{+0.12}$ & $0.30_{-0.04}^{+0.00}$ & $0.30_{-0.00}^{+0.03}$ & $0.30000$\\
$\sigma_{8}$ & $0.87_{-0.07}^{+0.10}$ & $0.92_{-0.21}^{+0.01}$ & $0.92_{-0.11}^{+0.04}$ & $0.82_{-0.14}^{+0.10}$ & $0.87_{-0.07}^{+0.05}$ & $0.84_{-0.03}^{+0.11}$ & $0.80000$\\
$S_{8}$ & $0.85_{-0.09}^{+0.18}$ & $0.84_{-0.13}^{+0.01}$ & $0.91_{-0.11}^{+0.06}$ & $0.72_{-0.27}^{+0.13}$ & $0.80_{-0.02}^{+0.12}$ & $0.89_{-0.09}^{+0.06}$ & $0.80000$\\
$A_{\textrm{SN1}}$ & $1.00_{-0.48}^{+0.96}$ & $2.56_{-0.45}^{+0.48}$ & $0.38_{-0.12}^{+0.75}$ & $2.10_{-0.62}^{+0.91}$ & $1.00_{-0.05}^{+0.13}$ & $0.44_{-0.18}^{+0.56}$ & $1.00000$\\
$A_{\textrm{SN2}}$ & $1.22_{-0.22}^{+0.26}$ & $0.59_{-0.04}^{+0.05}$ & $0.68_{-0.16}^{+0.49}$ & $0.64_{-0.13}^{+0.20}$ & $1.00_{-0.03}^{+0.44}$ & $0.68_{-0.16}^{+0.49}$ & $1.00000$\\
$A_{\textrm{AGN1}}$ & $1.00_{-0.63}^{+1.30}$ & $1.26_{-0.81}^{+0.78}$ & $1.00_{-0.26}^{+2.39}$ & $0.58_{-0.30}^{+0.93}$ & $1.71_{-0.71}^{+1.85}$ & $0.86_{-0.41}^{+1.84}$ & $1.00000$\\
$A_{\textrm{AGN2}}$ & $1.03_{-0.38}^{+0.84}$ & $0.95_{-0.01}^{+0.51}$ & $1.00_{-0.20}^{+0.68}$ & $0.95_{-0.36}^{+0.90}$ & $1.74_{-0.87}^{+0.26}$ & $1.18_{-0.38}^{+0.73}$ & $1.00000$\\
$\chi^{2}$ & $169_{-100}^{+114}$ & $77_{-27}^{+30}$ & $232_{-101}^{+89}$ & $65_{-34}^{+42}$ & $946_{-329}^{+405}$ & $539_{-199}^{+144}$ & $-$\\
$\tilde{\chi}^{2}$ & $0.79_{-0.18}^{+0.28}$ & $0.49_{-0.15}^{+0.14}$ & $0.68_{-0.13}^{+0.18}$ & $0.50_{-0.17}^{+0.32}$ & $4.33_{-0.96}^{+1.13}$ & $1.58_{-0.31}^{+0.28}$ & $-$\\
\hline
\end{tabular}
\label{tab:consistency_test_fiducial}
\end{table*}

\begin{table*}
\centering
\renewcommand*{\arraystretch}{1.50}
\caption{Same as Table \ref{tab:consistency_test_fiducial}, but considering a mock observational sample taken from the IllustrisTNG simulation `LH-698', which is considered as ground-truth.}

\begin{tabular}{cccccccc}
\hline
\hline
Parameter & $R_{*,1/2}$-$M_{*}$ & $f_{\textrm{DM}}(<R_{*,1/2})$-$M_{*}$ & $M_{\textrm{DM},1/2}$-$M_{*}$ & $M_{\textrm{tot}}$-$M_{*}$ & cumulative & cumulative alt. & ground-truth\\
\hline 
$\Omega_{\textrm{m}}$ & $0.23_{-0.02}^{+0.07}$ & $0.21_{-0.01}^{+0.01}$ & $0.24_{-0.00}^{+0.02}$ & $0.36_{-0.22}^{+0.11}$ & $0.27_{-0.00}^{+0.02}$ & $0.24_{-0.00}^{+0.02}$ & $0.26740$\\
$\sigma_{8}$ & $0.88_{-0.08}^{+0.05}$ & $0.97_{-0.10}^{+0.00}$ & $0.90_{-0.07}^{+0.00}$ & $0.94_{-0.12}^{+0.04}$ & $0.83_{-0.00}^{+0.00}$ & $0.90_{-0.07}^{+0.00}$ & $0.82580$\\
$S_{8}$ & $0.78_{-0.04}^{+0.03}$ & $0.80_{-0.11}^{+0.00}$ & $0.80_{-0.02}^{+0.01}$ & $0.97_{-0.30}^{+0.20}$ & $0.78_{-0.00}^{+0.02}$ & $0.81_{-0.03}^{+0.00}$ & $0.77964$\\
$A_{\textrm{SN1}}$ & $0.72_{-0.42}^{+0.11}$ & $2.41_{-0.91}^{+1.28}$ & $0.49_{-0.01}^{+0.00}$ & $0.32_{-0.04}^{+0.95}$ & $0.48_{-0.00}^{+0.00}$ & $0.49_{-0.02}^{+0.00}$ & $0.47500$\\
$A_{\textrm{SN2}}$ & $1.60_{-0.48}^{+0.02}$ & $0.56_{-0.05}^{+0.05}$ & $0.92_{-0.35}^{+0.33}$ & $0.67_{-0.13}^{+0.33}$ & $1.24_{-0.24}^{+0.00}$ & $0.56_{-0.00}^{+0.68}$ & $1.24401$\\
$A_{\textrm{AGN1}}$ & $1.00_{-0.55}^{+1.88}$ & $0.93_{-0.52}^{+0.58}$ & $1.00_{-0.73}^{+1.53}$ & $0.73_{-0.43}^{+1.21}$ & $2.53_{-1.47}^{+0.00}$ & $0.27_{-0.00}^{+2.26}$ & $2.52801$\\
$A_{\textrm{AGN2}}$ & $1.20_{-0.27}^{+0.60}$ & $1.27_{-0.68}^{+0.54}$ & $1.00_{-0.09}^{+0.79}$ & $1.04_{-0.24}^{+0.83}$ & $1.79_{-0.79}^{+0.00}$ & $0.91_{-0.00}^{+0.88}$ & $1.79378$\\
$\chi^{2}$ & $228_{-81}^{+226}$ & $83_{-39}^{+29}$ & $273_{-88}^{+133}$ & $180_{-118}^{+65}$ & $1222_{-257}^{+466}$ & $589_{-163}^{+205}$ & $-$\\
$\tilde{\chi}^{2}$ & $1.23_{-0.38}^{+0.29}$ & $0.57_{-0.12}^{+0.17}$ & $0.94_{-0.21}^{+0.40}$ & $0.59_{-0.14}^{+0.24}$ & $4.86_{-0.85}^{+1.58}$ & $1.97_{-0.44}^{+0.76}$ & $-$\\
\hline
\end{tabular}
\label{tab:consistency_test_bestfit}
\end{table*}

\begin{table*}
\centering
\renewcommand*{\arraystretch}{1.50}
\caption{Same as Table \ref{tab:consistency_test_fiducial}, but considering a mock observational sample taken from the IllustrisTNG simulation `LH-277', which is considered as ground truth.}

\begin{tabular}{cccccccc}
\hline
\hline
Parameter & $R_{*,1/2}$-$M_{*}$ & $f_{\textrm{DM}}(<R_{*,1/2})$-$M_{*}$ & $M_{\textrm{DM},1/2}$-$M_{*}$ & $M_{\textrm{tot}}$-$M_{*}$ & cumulative & cumulative alt. & ground-truth\\
\hline 
$\Omega_{\textrm{m}}$ & $0.31_{-0.06}^{+0.06}$ & $0.30_{-0.04}^{+0.02}$ & $0.26_{-0.05}^{+0.05}$ & $0.35_{-0.21}^{+0.12}$ & $0.33_{-0.03}^{+0.02}$ & $0.30_{-0.03}^{+0.03}$ & $0.33300$\\
$\sigma_{8}$ & $0.89_{-0.09}^{+0.07}$ & $0.90_{-0.17}^{+0.02}$ & $0.81_{-0.13}^{+0.16}$ & $0.97_{-0.21}^{+0.01}$ & $0.99_{-0.14}^{+0.00}$ & $0.95_{-0.13}^{+0.04}$ & $0.87660$\\
$S_{8}$ & $0.89_{-0.13}^{+0.16}$ & $0.85_{-0.13}^{+0.08}$ & $0.76_{-0.15}^{+0.20}$ & $1.06_{-0.55}^{+0.16}$ & $0.99_{-0.10}^{+0.06}$ & $0.95_{-0.16}^{+0.09}$ & $0.92356$\\
$A_{\textrm{SN1}}$ & $0.64_{-0.28}^{+1.02}$ & $2.60_{-0.96}^{+0.43}$ & $1.42_{-0.93}^{+0.66}$ & $0.81_{-0.52}^{+0.52}$ & $0.64_{-0.28}^{+0.31}$ & $0.68_{-0.34}^{+0.70}$ & $0.45313$\\
$A_{\textrm{SN2}}$ & $1.46_{-0.35}^{+0.26}$ & $0.55_{-0.02}^{+0.04}$ & $0.65_{-0.09}^{+0.51}$ & $0.69_{-0.16}^{+0.87}$ & $1.18_{-0.30}^{+0.26}$ & $0.81_{-0.30}^{+0.39}$ & $1.94127$\\
$A_{\textrm{AGN1}}$ & $1.78_{-1.41}^{+0.79}$ & $0.54_{-0.24}^{+1.34}$ & $0.94_{-0.51}^{+1.15}$ & $1.12_{-0.76}^{+1.07}$ & $1.71_{-1.34}^{+0.38}$ & $0.94_{-0.52}^{+1.16}$ & $2.17046$\\
$A_{\textrm{AGN2}}$ & $1.31_{-0.62}^{+0.58}$ & $1.24_{-0.51}^{+0.41}$ & $1.19_{-0.64}^{+0.32}$ & $1.01_{-0.25}^{+0.53}$ & $1.38_{-0.69}^{+0.36}$ & $1.11_{-0.43}^{+0.56}$ & $0.89689$\\
$\chi^{2}$ & $89_{-40}^{+41}$ & $39_{-16}^{+22}$ & $82_{-34}^{+77}$ & $80_{-51}^{+61}$ & $654_{-213}^{+425}$ & $272_{-118}^{+155}$ & $-$\\
$\tilde{\chi}^{2}$ & $0.47_{-0.11}^{+0.12}$ & $0.24_{-0.08}^{+0.10}$ & $0.40_{-0.09}^{+0.16}$ & $0.40_{-0.19}^{+0.22}$ & $2.76_{-0.70}^{+0.73}$ & $0.98_{-0.24}^{+0.31}$ & $-$\\
\hline
\end{tabular}
\label{tab:consistency_test_LH_277}
\end{table*}

\begin{table*}
\centering
\renewcommand*{\arraystretch}{1.50}
\caption{Same as Table \ref{tab:consistency_test_fiducial}, but considering a mock observational sample taken from the IllustrisTNG simulation `LH-25', which is considered as ground truth.}

\begin{tabular}{cccccccc}
\hline
\hline
Parameter & $R_{*,1/2}$-$M_{*}$ & $f_{\textrm{DM}}(<R_{*,1/2})$-$M_{*}$ & $M_{\textrm{DM},1/2}$-$M_{*}$ & $M_{\textrm{tot}}$-$M_{*}$ & cumulative & cumulative alt. & ground-truth\\
\hline 
$\Omega_{\textrm{m}}$ & $0.45_{-0.09}^{+0.01}$ & $0.16_{-0.02}^{+0.02}$ & $0.20_{-0.02}^{+0.03}$ & $0.33_{-0.23}^{+0.14}$ & $0.14_{-0.01}^{+0.02}$ & $0.18_{-0.02}^{+0.01}$ & $0.13260$\\
$\sigma_{8}$ & $0.87_{-0.07}^{+0.01}$ & $0.90_{-0.21}^{+0.04}$ & $0.81_{-0.17}^{+0.12}$ & $0.92_{-0.12}^{+0.05}$ & $0.88_{-0.18}^{+0.08}$ & $0.72_{-0.10}^{+0.22}$ & $0.79380$\\
$S_{8}$ & $1.08_{-0.20}^{+0.00}$ & $0.67_{-0.15}^{+0.00}$ & $0.68_{-0.14}^{+0.09}$ & $0.89_{-0.43}^{+0.25}$ & $0.62_{-0.11}^{+0.05}$ & $0.58_{-0.08}^{+0.09}$ & $0.52774$\\
$A_{\textrm{SN1}}$ & $3.19_{-1.17}^{+0.62}$ & $0.30_{-0.04}^{+0.18}$ & $0.38_{-0.10}^{+0.26}$ & $0.36_{-0.08}^{+0.64}$ & $0.80_{-0.19}^{+0.54}$ & $0.30_{-0.02}^{+0.33}$ & $1.05263$\\
$A_{\textrm{SN2}}$ & $1.50_{-0.35}^{+0.11}$ & $1.62_{-0.03}^{+0.23}$ & $1.54_{-0.49}^{+0.28}$ & $1.00_{-0.33}^{+0.57}$ & $0.62_{-0.11}^{+0.57}$ & $1.62_{-0.99}^{+0.22}$ & $0.54450$\\
$A_{\textrm{AGN1}}$ & $0.76_{-0.06}^{+3.10}$ & $2.71_{-2.40}^{+0.99}$ & $1.29_{-0.94}^{+1.80}$ & $1.26_{-0.89}^{+1.27}$ & $1.38_{-1.06}^{+1.35}$ & $0.80_{-0.39}^{+2.59}$ & $3.95040$\\
$A_{\textrm{AGN2}}$ & $1.22_{-0.54}^{+0.57}$ & $0.67_{-0.12}^{+0.81}$ & $1.21_{-0.58}^{+0.23}$ & $1.00_{-0.21}^{+0.31}$ & $1.09_{-0.47}^{+0.49}$ & $1.14_{-0.53}^{+0.45}$ & $0.56762$\\
$\chi^{2}$ & $9_{-6}^{+15}$ & $71_{-42}^{+62}$ & $44_{-25}^{+27}$ & $130_{-94}^{+189}$ & $543_{-290}^{+652}$ & $167_{-71}^{+180}$ & $-$\\
$\tilde{\chi}^{2}$ & $0.14_{-0.06}^{+0.09}$ & $0.34_{-0.16}^{+0.28}$ & $0.20_{-0.12}^{+0.10}$ & $0.60_{-0.40}^{+0.49}$ & $3.88_{-2.15}^{+2.84}$ & $0.84_{-0.32}^{+0.56}$ & $-$\\
\hline
\end{tabular}
\label{tab:consistency_test_LH_25}
\end{table*}


\bsp	
\label{lastpage}
\end{document}